\documentclass[journal]{IEEEtran}
\usepackage{amsmath,amsfonts}
\usepackage{algorithmic}
\usepackage{algorithm}
\usepackage{array}
\usepackage[caption=false,font=normalsize,labelfont=sf,textfont=sf]{subfig}
\usepackage{textcomp}
\usepackage{booktabs}
\usepackage{stfloats}
\usepackage{url}
\usepackage{verbatim}
\usepackage{graphicx}
\usepackage{cite}
\usepackage{rotating}
\usepackage{tabularray}
\usepackage{multirow}
\usepackage{multicol}
\usepackage{color}

\usepackage{soul}
\usepackage{color, xcolor}
\usepackage[pdfstartview=XYZ, bookmarks=true, colorlinks=true, linkcolor=blue, urlcolor=blue, citecolor=blue, pdftex, bookmarks=true, linktocpage=true,  hyperindex=true]{hyperref}

\begin{document}

\title{BDAN: Mitigating Temporal Difference Across Electrodes in Cross-Subject Motor Imagery Classification via Generative Bridging Domain}

\author{Zhige Chen, Rui Yang, \IEEEmembership{Senior Member, IEEE}, Mengjie Huang, \IEEEmembership{Member, IEEE}, Chengxuan Qin, Zidong Wang, \IEEEmembership{Fellow, IEEE}
\thanks{This research has been approved by University Ethics Committee of Xi'an Jiaotong-Liverpool University with proposal number EXT20-01-07 on March 31 2020, and is partially supported by: Jiangsu Provincial Qinglan Project, the Natural Science Foundation of the Jiangsu Higher Education Institutions of China (23KJB520038), Suzhou Science and Technology Programme (SYG202106), and the Research Enhancement Fund of XJTLU (REF-23-01-008).}
\thanks{Z.~Chen and C.~Qin are with School of Advanced Technology, Xi'an Jiaotong-Liverpool University, Suzhou, 215123, China, and also with School of Electrical Engineering, Electronics and Computer Science, University of Liverpool, Liverpool, L69 3BX, United Kingdom (e-mail: Zhige.Chen21@alumni.xjtlu.edu.cn; Chengxuan.Qin21@student.xjtlu.edu.cn).}
\thanks{R.~Yang is with School of Advanced Technology, Xi'an Jiaotong-Liverpool University, Suzhou, 215123, China (e-mail: R.Yang@xjtlu.edu.cn).}
\thanks{M.~Huang is with Design School, Xi'an Jiaotong-Liverpool University, Suzhou, 215123, China (e-mail: Mengjie.Huang@xjtlu.edu.cn).}
\thanks{Z.~Wang is with the Department of Computer Science, Brunel University London, Uxbridge, Middlesex, UB8 3PH, United Kingdom (e-mail: Zidong.Wang@brunel.ac.uk).}
\thanks{Corresponding authors: R.~Yang and M.~Huang.}}

\maketitle

\begin{abstract}
Because of ``the non-repeatability of the experiment settings and
conditions" and ``the variability of brain patterns among subjects", the data distributions across sessions and electrodes are different in cross-subject motor imagery (MI) studies, eventually reducing the performance of the classification model. Systematically summarised based on the existing studies, a novel temporal-electrode data distribution problem is investigated under both intra-subject and inter-subject scenarios in this paper. Based on the presented issue, a novel bridging domain adaptation network (BDAN) is proposed, aiming to minimise the data distribution difference across sessions in the aspect of the electrode, thus improving and enhancing model performance. In the proposed BDAN, deep features of all the EEG data are extracted via a specially designed spatial feature extractor. With the obtained spatio-temporal features, a special generative bridging domain is established, bridging the data from all the subjects across sessions. The difference across sessions and electrodes is then minimized using the customized bridging loss functions, and the known knowledge is automatically transferred through the constructed bridging domain. To show the effectiveness of the proposed BDAN, comparison experiments and ablation studies are conducted on a public EEG dataset. The overall comparison results demonstrate the superior performance of the proposed BDAN compared with the other advanced deep learning and domain adaptation methods.

\end{abstract}


\section{Introduction}
\label{Introduction}
Distinct activities and brain states within the mammalian brain induce varying patterns of cortical synchrony, leading to a diverse array of electrical potentials across the scalp over time. These spatio-temporal related scalp potentials are posited as a core mechanism for neuronal computation and can be effectively captured and recorded through electroencephalogram (EEG). With high precision in detecting and pinpointing signals from multi-channel electrodes, EEG has become a pivotal tool for recording and analyzing various brain activities, such as motor imagery (MI). EEG-based motor imagery (MI) represents the neural activities in the motor cortex responsible for the planning and execution of movement, establishing a communication bridge between humans and external devices. The EEG-based MI classification has thus been under the spotlight in recent years.

MI signals consist of features from temporal, frequency, and spatial domains, significantly elevating the complexity in the extraction of relevant features. With the dramatic improvement of computational devices \cite{Zhang2023MultipleGraph}, deep learning (DL) models \cite{Chen2023SMA, Dong2023OptimalAttacks} have exhibited remarkable proficiency in multi-domain feature extraction, widely adopted in the processing and analysis of MI signals. Despite the aforementioned advantages of the DL models, the data shortages and variabilities still pose great challenges to the performance and generalizability of the classification model. Cross-subject MI classification approaches have been developed to tackle these challenges, maintaining the learning performance of the classification model by transferring known knowledge \cite{Liao2023TransferLearning} from the source subject, ultimately accomplishing unsupervised classification of the target subject with limited data. 

Based on the number of subjects from the source and target domains, cross-subject MI classification can be broadly categorized into three categories: single-source to single-target (STS), multi-source to single-target (MTS), and multi-source to multi-target (MTM). Among the scope of these classification tasks, the STS problem stands out as notably arduous attributed to the paucity of available training data. Moreover, the substantial heterogeneity across subjects results in the data distribution difference of inter-subject, reducing the generalizability of the classification model, especially in the STS task. To address the cross-subject MI classification problem, transfer learning methods such as data alignment and domain adaptation methods have been proposed and widely adopted in cross-subject scenarios, aiming to reduce the data distribution difference of inter-subject and improve the performance of the classification model \cite{Chen2023EDAN}.

The existing cross-subject MI classification studies are presented in Table \ref{Cross-subject MI studies}, consisting of transfer learning (TL) methods using data alignment and domain adaptation algorithms focusing on multiple spaces/domains. As shown in Table \ref{Cross-subject MI studies}, the existing TL methods mainly focus on the global space/domain of MI signals, while a few studies conduct temporal and spatial alignment/adaptation. Furthermore, the existing studies in the temporal space/domain mainly focus on the global data distribution difference across sessions, but no literature investigates the temporal data distribution difference across electrodes under both intra- and inter-subject scenarios. Because of the spatio-temporal variability of MI data, such temporal-electrode data distribution difference problems will appear in all MI experiments, especially emerging in the cross-subject MI scenarios.

\begin{table}[t]
\centering
\setlength{\tabcolsep}{4pt}
\caption{Summary of the Existing MI Classification Studies}
\label{Cross-subject MI studies}
\begin{tabular}{cccc} 
\hline\hline
\rule{0pt}{9pt}
\textbf{TL methods}  & \textbf{MTM} & \multicolumn{1}{c}{\textbf{MTS}}  &\textbf{STS} \\ [0.6ex]
\hline
\rule{0pt}{13pt}
\begin{tabular}[c]{@{}c@{}}\rule{0pt}{7pt}Global space \\ data alignment\end{tabular}  & Nil  & \multicolumn{1}{c}{\begin{tabular}[c]{@{}c@{}}\cite{Cheng2017CSP, Cai2022MIDecoding, Liang2020MultiFusionTL, Zheng2020EEGAcrossSession}\end{tabular}}  & Nil  \\ [2ex]
\begin{tabular}[c]{@{}c@{}}Temporal space \\data alignment\end{tabular}  & \cite{He2019EA} & \cite{Tang2023RA}  & Nil  
\\ [2ex] 
\begin{tabular}[c]{@{}c@{}}Spatial space \\data alignment\end{tabular}  & \cite{He2019EA} & \cite{Tang2023RA}  & Nil  
\\ [2ex] 
\hline
\rule{0pt}{13pt}
\begin{tabular}[c]{@{}c@{}}\rule{0pt}{10pt}Global domain \\ adaptation\end{tabular} & \cite{Hu2021FederatedTL, Ozdenizci2020AdversarialEEG}  & \cite{Hong2021DJDA, She2023DAWD, Zhao2020JDA, Zhong2023DDAF}  & \cite{Zhao2020NonstationaryTL} \\ [2.5ex]
\begin{tabular}[c]{@{}c@{}}Temporal domain \\adaptation\end{tabular}  & Nil & \cite{Azab2019WeightedTL, Zhang2023CrossSessionMI}  & \cite{Chen2022STSTL, Liu2024CrossSessionMI}    
\\ [2.5ex]
\begin{tabular}[c]{@{}c@{}}Spatial domain \\adaptation\end{tabular} & \cite{Li2023GDA, Perez2022EEGSYM} & \cite{Chen2021MultiAttention, Shi2022ChannelSelection} & Nil  \\ [2.5ex]
\begin{tabular}[c]{@{}c@{}}Spatio-temporal domain \\adaptation\end{tabular} & Nil & Nil & Nil  
\\[1.5ex] 
\hline\hline     
\end{tabular}
\end{table}

\begin{figure}[t]
      \centering
      \includegraphics[width=1\linewidth]{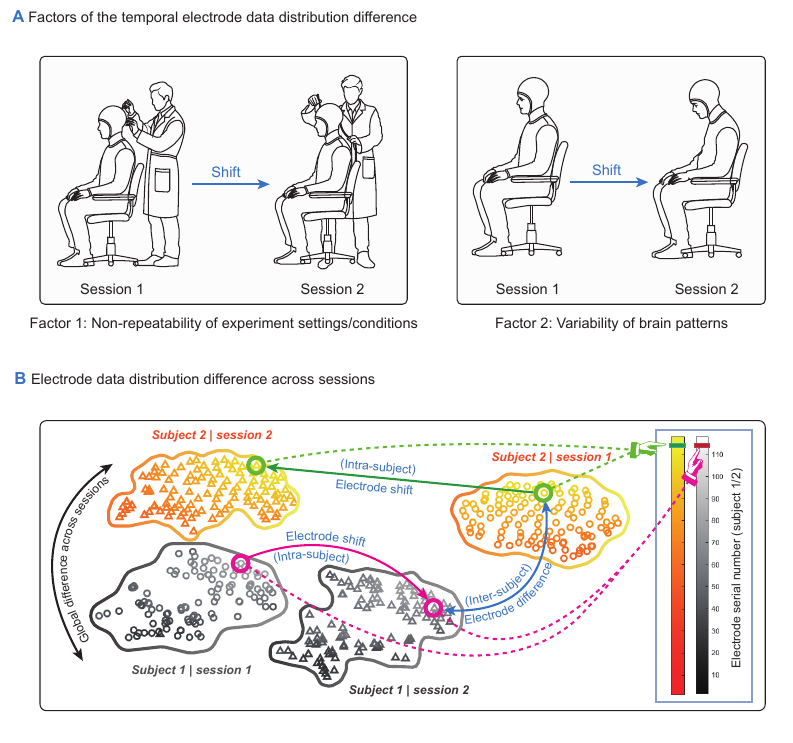}
      \caption{Schematic of the proposed temporal-electrode data distribution difference problem. \textbf{A}, Factors of the temporal-electrode data distribution difference problem, where the two main factors are illustrated diagrammatically. \textbf{B}, Electrode data distribution difference across sessions, where the proposed temporal-electrode data distribution difference is diagrammatically presented.}
      \label{Factors TEDA}
\end{figure}

A schematic of the investigated temporal-electrode data distribution difference problem is shown in Fig. \ref{Factors TEDA}, consisting of a diagrammatic illustration of factors (Fig. \ref{Factors TEDA}A) and a schematic data distribution across sessions and electrodes (Fig. \ref{Factors TEDA}B). According to the literature review, such a non-negligible temporal-electrode data distribution difference problem in MI-related studies is mainly caused by two factors:

\textit{\textbf{1) Non-repeatability of experiment settings and conditions.}} In the MI experiment, electrodes are manually positioned at different montages on the brain scalp following different International electrode systems such as 10/5, 10/10, and 10/20 systems. With the recorded multi-channel EEG signals, the different cognitive activities of the subjects are eventually analysed and classified. However, as shown in ``Factor 1" in Fig. \ref{Factors TEDA}A, the non-repeatability of manual positioning and calibration \cite{Qin2023SVGA}, as well as the different operating habits of different operators \cite{Liu2023EEGLocation}, inevitably result in the variability of experimental settings across different sessions. Furthermore, the external environmental changes in the experiment, such as variabilities in the conductive medium \cite{Li2023ElectrodeRes}, the contact condition between electrode and skin \cite{Qin2023SVGA, Qin2023BrainSourceLocation}, and the temperature, will all evoke changes to the electrode impedance, thus impairing the quality of EEG data and resulting in temporal-electrode data distribution difference. Experiment settings and conditions proved to be elusive with exact replication in each experiment duration, inducing temporal-electrode data distribution differences and ultimately degrading the model's classification performance.

\textit{\textbf{2) Variability of brain patterns among subjects.}} MI experiments generally incorporate multiple subjects, each of whom possesses distinct brain structures and functions (brain patterns), including the morphological difference, the neural conduction velocities \cite{Xu2023BrainDynamics}, and the connections between neurons \cite{Staresina2023NeuroCommunication}. In MI experiments, as shown in ``Factor 2" in Fig. \ref{Factors TEDA}A, with the time-consuming data acquisition, the concentration, emotional state, and fatigue level of the subject \cite{Camacho2023BrainEmotion} will dramatically affect the brain's activity pattern and consequently compromise the data recorded by each electrode. In addition, subjects will gradually acclimatize to the experimental settings and adapt their response pattern over continuous and repeated tasks \cite{Reber2023NeuralAdaptation}, thus resulting in changes in the duration and way of connections across brain regions, ultimately leading to temporal-electrode data distribution differences. In conclusion, because of individual variability, fatigue degrees, mental conditions, and brain plasticity, brain patterns will shift incessantly among subjects and across sessions in MI experiments, thus leading to the temporal-electrode data distribution difference.

Because of the factors mentioned in aspects \textit{1)} and \textit{2)}, the electrode data distribution will shift across sessions under both intra- and inter-subject scenarios. As shown in Fig. \ref{Factors TEDA}B, represented by the circles and triangles respectively, the data distribution of the same electrode (with the same gradient colour indicated by the colour bar) will shift across sessions, bringing the global data distribution difference across sessions under both intra- and inter-subject scenarios, thus increasing the data distribution difference among subjects. However, the temporal-electrode data distribution caused by the two factors cannot be fully eliminated in cross-subject motor imagery (MI) experiments, ultimately affecting the robustness and performance of STS MI classification models. To the best of the authors' knowledge, no previous work has addressed the spatio-temporal data distribution difference problem across both sessions and electrodes, and there is a need to reduce such temporal-electrode data distribution differences and improve the existing STS MI classification algorithms. To address such a problem, we propose a novel bridging domain adaptation network (BDAN) consisting of a specially designed spatial feature extractor and bridging domain adaptation, aiming to eliminate the proposed problem under both intra- and inter-subject scenarios.

\begin{figure*}[t]
      \centering
      \includegraphics[width=1\linewidth]{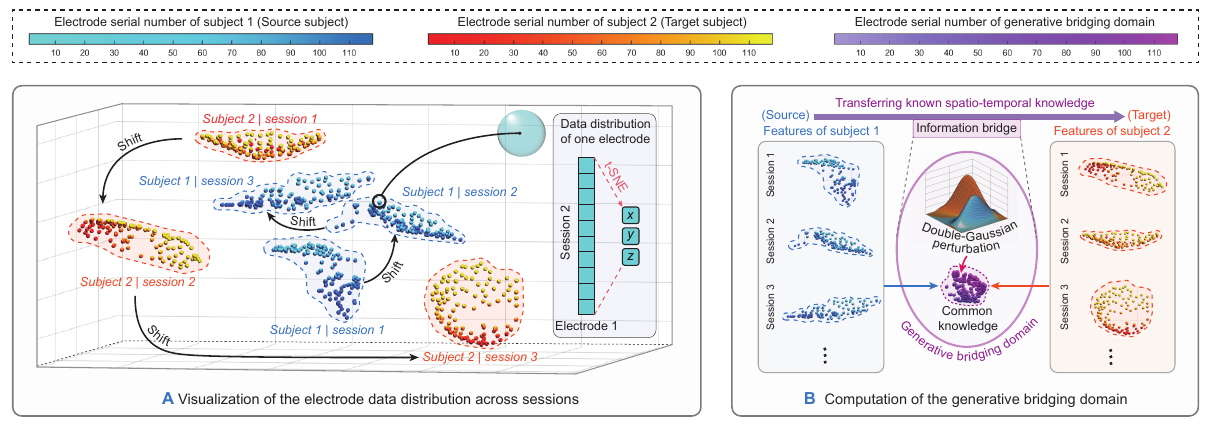}
      \caption{Visualization of temporal-electrode data distribution problem and solutions. \textbf{A}, Electrode data distribution across sessions from different subjects, where each electrode ball represents the data distribution of one electrode, the colour of the electrode ball indicates the corresponding subject and electrode, and the arrows denote the data distribution differences across sessions. \textbf{B}, Computation of the generative bridging domain, where the bridging domain is computed via all the features from the source and target subjects.}
      \label{Visualization_of_problem}
\end{figure*}

The key contributions of this paper are summarised as follows: \par
1) Investigation of the temporal-electrode data distribution problem, considering the spatio-temporal data distribution difference problem across both sessions and electrodes under both intra- and inter-subject scenarios;

2) Introduction of a novel domain adaptation algorithm called BDAN, where the bridging domain adaptation is conducted on the extracted spatio-temporal features to effectively reduce the data distribution differences across both sessions and electrodes;

3) Establishment of a generative bridging domain, mitigating the temporal-electrode data distribution difference of inter-subjects with no need of electrodes and session labels of the input EEG data and significantly reducing the complexity of the domain adaptation across sessions;

4) Construction of the specially designed matrix arithmetic operation of the established bridging domain loss functions with low computational overhead, where the distances across electrodes and sessions are computed via dot product after matrix transformation instead of traversing all electrodes and samples;

5) Successful comparison experiments and ablation studies conducted on two publicly available brain-computer interface (BCI) competition datasets, demonstrating the superior performance and robustness of the proposed BDAN method over existing approaches.

The remaining sections of this paper are organized as follows. Section \ref{Methodology} outlines the temporal-electrode data distribution difference problem and presents the proposed BDAN algorithm step by step. The experiments and results of the proposed method are presented and analyzed in Section \ref{Experiment and Result Analysis} through comparison and visualization analyses. Finally, Section \ref{Conclusion} concludes the paper and discusses potential future work. \par

\section{Methodology}
\label{Methodology}

To deal with the proposed temporal-electrode data distribution difference problem, we propose BDAN, a bridging domain adaptation network consisting of a specially designed spatial feature extractor and bridging domain adaptation. Our goal is to minimize the spatio-temporal data distribution difference of inter-subject in the aspect of the bridging domain. Instead of directly evaluating the difference across sessions and electrodes, the temporal-electrode data distribution difference is calculated and adaptively minimized via a generative bridging domain obtained by deep features, bridging and transferring the knowledge of all the sessions from the source and target domains. In this section, the investigated temporal-electrode data distribution problem is visualized and analyzed in detail, and the proposed BDAN approach is presented in four different aspects.

\begin{figure*}[t]
      \centering
      \includegraphics[width=1\linewidth]{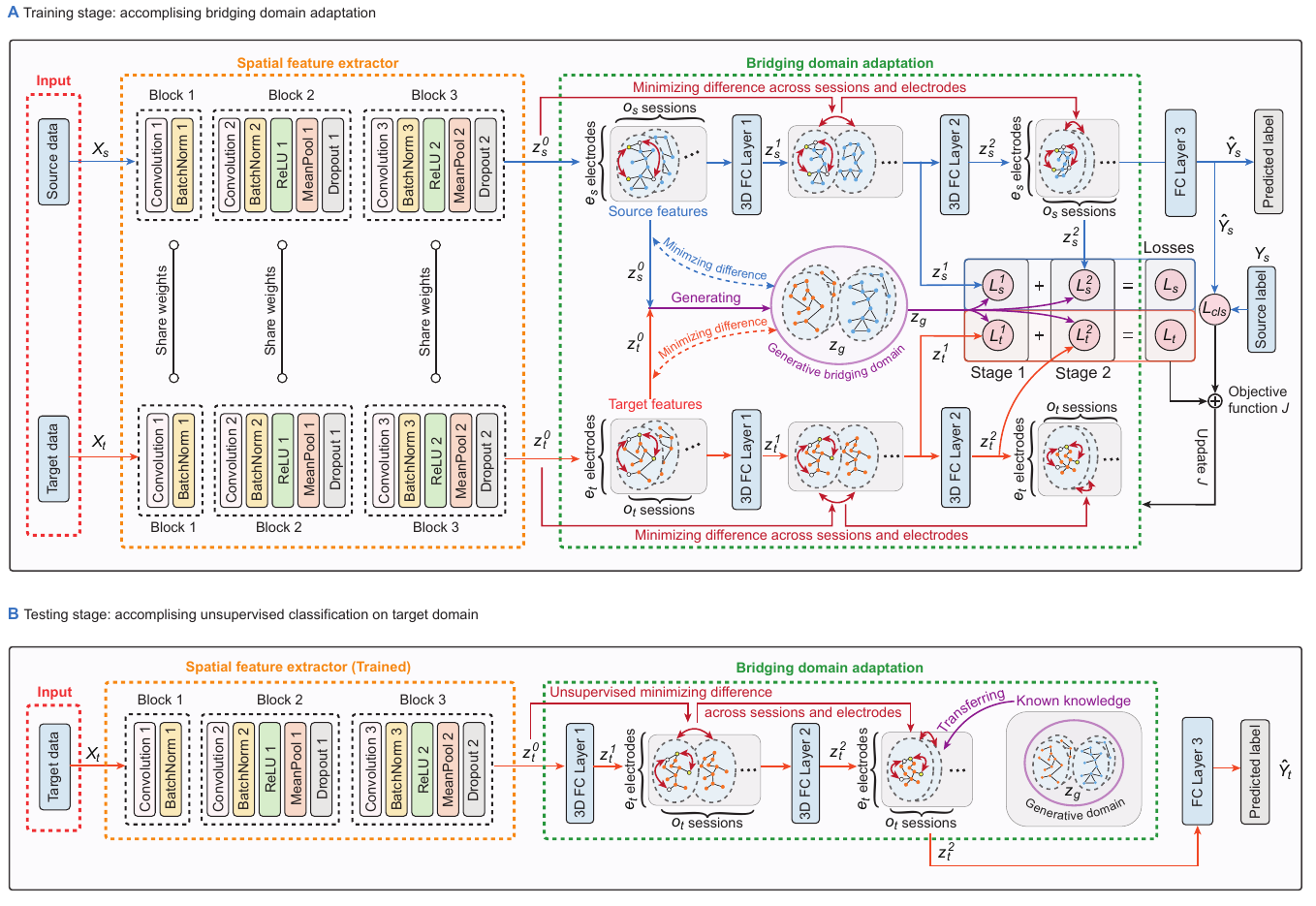}
      \caption{Bridging domain adaptation network (BDAN), where the blue and orange arrows represent the data flows from the source and target domains respectively, and the purple arrow represents the data flow of the generative bridging domain. \textbf{A}, Training stage of the BDAN, where bridging domain adaptation is adopted between source and target data, minimising the temporal-electrode data distribution difference and transferring the known spatio-temporal knowledge from the source domain to the target domain. \textbf{B}, Testing stage of the BDAN, where unsupervised classification is adopted on the target domain via the trained model, automatically reducing the time-related electrode data distribution difference and improving the classification performance.}
      \label{Framework}
\end{figure*}

\subsection{Visualization of temporal-electrode data distribution}
The data distribution shifted with time of each electrode from different subjects is shown in Fig. \ref{Visualization_of_problem}A. Each electrode ball represents the data distribution of the corresponding electrode (with electrode serial number indicated by the colour bar) obtained by the t-SNE method, manifesting the electrode data distribution difference shifted with time under both intra- and inter-subject scenarios. As shown in Fig. \ref{Visualization_of_problem}A, electrode balls from subjects 1 and 2 are coloured with the graduated colour blue and orange, respectively, and each subject consists of the data from multiple consecutive experiment sessions. The proposed temporal-electrode data distribution difference problem can be summarised in two aspects by observing the distribution of electrode balls in Fig. \ref{Visualization_of_problem}A:

1) When it comes to intra-subject scenarios (represented by the electrode balls with the same graduated colors): as shown by the black arrows in Fig. \ref{Visualization_of_problem}A, the locations of the same electrode balls of each subject shifted across sessions, in-coordinately deviating from the location of the initial experiment session; 

2) When it comes to inter-subject scenarios (represented by the electrode balls with different graduated colors): the locations and distances of the same electrode ball vary across sessions of inter-subject, indicating the significant variability of brain patterns among subjects.

The shift of each electrode ball contributes to the global shift of MI data across sessions, thus leading to the global discrepancy in locations and orientations across sessions under both intra- and inter-subject scenarios. Therefore, the observations in aspects 1) and 2) illustrate the proposed temporal-electrode data distribution problem under both intra- and inter-subject scenarios. The basic constituent elements within the EEG data are the electrode channels, and because of the aforementioned factors in Section \ref{Introduction}, the data distribution difference across sessions and electrodes can not be eliminated in EEG experiment, thus giving rise to the global data distribution difference of inter-subject. 

Therefore, to increase the performance and transferability in STS MI classification task, the data difference across sessions and electrodes needs to be minimized. In cross-subject MI classification task, EEG data consists of multiple experiment sessions from multiple subjects, instead of directly evaluating the difference across sessions and electrodes of intra- and inter-subject, we calculate and adaptively minimize the temporal-electrode data distribution difference via a special generative domain (called bridging domain). As shown in Fig. \ref{Visualization_of_problem}B, the generative bridging domain obtained by extracted features consists of the common knowledge from all the subjects and sessions, serving as an information transformation bridge between the source and target domains.

\subsection{Spatial feature extractor}
\label{Spatial feature extractor}

The proposed BDAN regards MI data from two subjects as inputs in the training stage (as shown in red dotted box ``Input'' in Fig. \ref{Framework}A), accomplishing unsupervised model training of the target domain by transferring the learnt knowledge from source data and labels. Let $X_s$ and $X_t$ represent the MI data from source and target domains respectively, both the source and target data are then loaded into the specially designed spatial feature extractor consisting of three functional blocks (as shown in orange dotted box in Fig. \ref{Framework}A), obtaining deep spatial features in the aspect of the electrode. The functions of the three blocks are presented as follows:
\begin{itemize}
    \item \textbf{\textit{Block 1}} for primary temporal convolution: temporal convolution is adopted on the input EEG data to extract temporal features, expanding the depth of the classification model;
    \item \textbf{\textit{Block 2}} for adjacent-electrode spatial convolution: spatial convolution is adopted across adjacent electrodes to extract spatial information, maintain the topology of the input EEG data;
    \item \textbf{\textit{Block 3}} for deep temporal convolution: temporal convolution is adopted on the features extracted by \textbf{\textit{Block 2}} to learn informative features.
\end{itemize}

\begin{table}[t]
\renewcommand\arraystretch{1.3}
\footnotesize
\centering
\caption{Parameters of the Spatial Feature Extractor}
\label{Parameters BDAN}
\begin{tabular}{p{2cm}|p{6cm}} 
\hline\hline
\textbf{Name} & \textbf{Layers and Options}  \\ [0.5ex]
\hline
\multicolumn{2}{c}{\textbf{``\textit{Block 1}" for primary temporal convolution}} \\ [0.5ex]
\hline
\textbf{Convolution 1} & \begin{tabular}[c]{@{}p{6cm}@{}}\textit{Convolutional layer} \\
                        Kernel size = (1, 25), Filters = 8, Padding = 'valid'\end{tabular}  \\ [2ex]
\textbf{BatchNorm 1}  & \begin{tabular}[c]{@{}p{6cm}@{}}\textit{Batch normalization} \\
                        Channels = 8, Momentum = 0.9, Epsilon = 1e-5\end{tabular}       \\ [2ex]
\hline
\multicolumn{2}{c}{\textbf{``\textit{Block 2}" for adjacent-electrode spatial convolution}} \\ 
\hline
\textbf{Convolution 2} & \begin{tabular}[c]{@{}p{6cm}@{}}\textit{Convolutional layer} \\
                        Kernel size = (3, 1), Filters = 16, Padding = 'same'\end{tabular}  \\ [2ex]
\textbf{BatchNorm 2} & \begin{tabular}[c]{@{}p{6cm}@{}}\textit{Batch normalization} \\
                        Channels = 16, Momentum = 0.9, Epsilon = 1e-5\end{tabular}       \\ [2ex]
\textbf{RELU 1} & \begin{tabular}[c]{@{}p{6cm}@{}}\textit{RELU layer} \\
                  Inplace = 'False'\end{tabular}  
                  \\ [2ex]
\textbf{MeanPool 1} & \begin{tabular}[c]{@{}p{6cm}@{}}\textit{Mean pooling layer} \\
                    Kernel size = (1, 5), Stride = (1, 5)\end{tabular}               \\  [2ex]
\textbf{Dropout 1} & \begin{tabular}[c]{@{}p{6cm}@{}}\textit{Dropout layer} \\
                    Dropout ratio = 0.1\end{tabular}                                          \\  [2ex]
\hline
\multicolumn{2}{c}{\textbf{``\textit{Block 3}" for deep temporal convolution}} \\ 
\hline
\textbf{Convolution 3}  & \begin{tabular}[c]{@{}p{6cm}@{}}\textit{Convolutional layer} \\
                            Kernel size = (1,10), Filters = 32, Padding = 'valid'\end{tabular}   \\ [2ex]
\textbf{BatchNorm 3}     & \begin{tabular}[c]{@{}p{6cm}@{}}\textit{Batch normalization} \\
                            Channels = 32, Momentum = 0.9, Epsilon = 1e-5\end{tabular}         \\ [2ex]
\textbf{RELU 2}     & \begin{tabular}[c]{@{}p{6cm}@{}}\textit{RELU layer} \\
                    Inplace = 'False'\end{tabular}  \\ [2ex]
\textbf{MeanPool 2} & \begin{tabular}[c]{@{}p{6cm}@{}}\textit{Mean pooling layer} \\
                    Kernel size = (1, 5), Stride = (1, 5)\end{tabular}                   \\ [2ex]
\textbf{Dropout 2} & \begin{tabular}[c]{@{}p{6cm}@{}}\textit{Dropout layer} \\
                    Dropout ratio = 0.1\end{tabular}                                          \\ [2ex]
\textbf{3D FC Layer 1}  & \textit{3D Linear fully-connected layer}\\ [2ex]
\textbf{3D FC Layer 2}  & \textit{3D Linear fully-connected layer}\\ [2ex]
\textbf{2D FC Layer 1}  & \textit{2D Linear fully-connected layer}\\ [0.5ex]
\hline\hline
\end{tabular}
\end{table}

\subsection{Bridging domain adaptation}

With the deep features $z_s^0$ and $z_t^0$ extracted by the spatial feature extractor, the data distribution difference across sessions and electrodes are then computed and minimized via bridging domain adaptation (as shown in the green dotted box in Fig. \ref{Framework}A), automatically transferring the learnt knowledge from $z_s^0$ to $z_t^0$ through the generative bridging domain. The bridging domain adaptation is the core of the proposed BDAN, consisting of two ``adaptation stages" with four customized loss functions ($L_s^1$, $L_s^2$, $L_t^1$ and $L_t^2$ in Fig. \ref{Framework}A). In this paper, we use customised bridging loss functions to evaluate the data distribution difference, minimizing the difference with the optimization of the classification model. The bridging domain adaptation is the core of the proposed BDAN and is presented in the following three aspects:

\subsubsection{Establishment of the generative bridging domain} \label{Calculation of the generative domain}

As shown in Fig. \ref{Visualization_of_problem}, generated by the deep features from the source and target domains, the generative bridging domain consists of information from all the sessions and electrodes, indicating the common data distribution of all the EEG data. To generate the bridging domain, the global mean between the source and target domain samples $\overline{z}_{g}\left(e, p\right)$ is first computed by the extracted deep features $z_s^0 \in \mathbb{R}^{n_{s} \times f_{s} \times e_{s} \times p_{s}}$ and $z_t^0 \in \mathbb{R}^{n_{t} \times f_{t} \times e_{t} \times p_{t}}$:

\begin{equation}
\label{Equation mean generative domain}
\begin{aligned}
\overline{z}_{s}^{0}\left(e, p\right) &= \frac{1}{n_{s}} \frac{1}{f_{s}} \sum_{i=1}^{n_{s}} \sum_{f=1}^{f_{s}} z_{s}^{0}\left(i, f, e, p\right) \\
\overline{z}_{t}^{0}\left(e, p\right) &= \frac{1}{n_{t}} \frac{1}{f_{t}} \sum_{j=1}^{n_{t}} \sum_{f=1}^{f_{t}} z_{t}^{0}\left(j, f, e, p\right) \\
\overline{z}_{g}\left(e, p\right) &= \frac{n_{s}}{n_{s}+n_{t}} \overline{z}_{s}^{0}\left(e, p\right) + \frac{n_{t}}{n_{s}+n_{t}} \overline{z}_{t}^{0}\left(e, p\right)
\end{aligned}
\end{equation}
where $\overline{z}_{s}^{0}\left(e, p\right)$ and $\overline{z}_{t}^{0}\left(e, p\right)$ denote the mean of all the samples of source and target domains in terms of electrode, $n_s$, $f_s$, $e_s$, and $p_s$ represent the samples, filters, electrodes, feature points of the source domain, $n_t$, $f_t$, $e_t$, and $p_t$ represent the samples, filters, electrodes, feature points of the target domain.

With the obtained global mean $\overline{z}_{g} \in \mathbb{R}^{n_{g} \times f_{g} \times e_{g} \times p_{g}}$, the samples of the bridging domain are generated with double-Gaussian perturbation, generating multiple bridging samples based on the standard deviations (STD) from the source and target domains. The samples of the generative bridging domain $z_g$ can be generated by:

\begin{equation}
\begin{gathered}
\begin{aligned}
z_g \in \mathbb{R}^{n_{g} \times f_{g} \times e_{g} \times p_{g}} &= \mathbf{1}_{n_g} \otimes \overline{z}_g + w_G \cdot \sigma_g \cdot \mathcal{N}_{\text{Gaussian}}
\end{aligned} \\
\begin{cases}
w_G = \sigma_s & \text{(For source domain)} \\
w_G = \sigma_t & \text{(For target domain)}
\end{cases}
\end{gathered}
\end{equation}
where $n_g$, $f_g$, $e_g$, and $p_g$ represent the generated samples, filters, electrodes, feature points of the bridging domain, $\otimes$ denotes the Kronecker product (indicating the global mean is duplicated $n_{g}$ times using an all-ones vector $\mathbf{1}_{n_g}$), $\mathcal{N}_{\text{Gaussian}}$ denotes the standard Gaussian noise with distribution $(0,1)$, $\sigma_g$ denotes the STD of the global mean $\overline{z}_{g}$, $\sigma_s$ and $\sigma_t$ denote the STD of the deep features from source and target domains.

With the generative bridging domain data $z_{g}$, the data distribution difference across sessions and electrodes under both intra- and inter-subject scenarios can be easily evaluated. In contrast to computing multiple differences among all the sessions and electrodes between subjects, the bridging domain will significantly reduce the complexity of the bridging domain adaptation (\textit{Remark 1}: the proposed BDAN accomplishes domain adaptation solely by aligning all the EEG data to samples in the bridging domain, with no need for one-to-one session adaptation).

\subsubsection{Tensor transformation operations}

\begin{figure*}[t]
      \centering
      \includegraphics[width=1\linewidth]{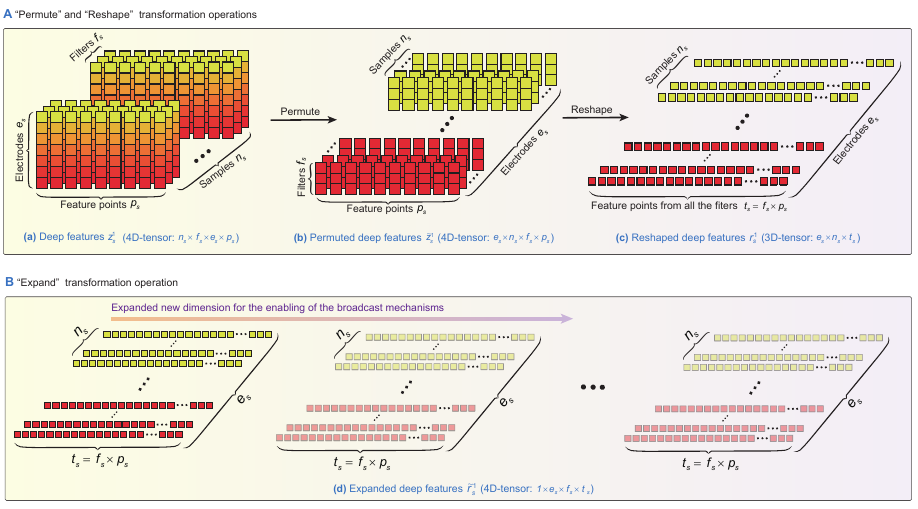}
      \caption{Transformation operations of deep features. \textbf{A}, ``Permute" and ``Reshape" transformation operations, where the 4D-tensor with shape $n_{s} \times f_{s} \times e_{s} \times p_{s}$ is transformed into 3D-tensor with shape $e_{s} \times n_{s} \times t_{s}$ via ``Permute" and ``Reshape" operations. \textbf{B}, ``Expand" transformation operation, where the 3D-tensor with shape $e_{s} \times n_{s} \times t_{s}$ is expanded into 4D-tensor with shape $1 \times e_{s} \times n_{s} \times t_{s}$.}
      \label{Operations}
\end{figure*}

In this paper, the specially designed bridging loss functions are the core of the bridging domain adaptation, with all the computing process among tensors performed in matrix arithmetic to expedite the computation speed via three different transformation operations: 1) ``Permute", 2) ``Reshape", and 3) ``Expand". 

The detailed illustrations of ``Permute", ``Reshape", and ``Expand" transformation operations of $z_{s}^{1}$ are shown in Fig. \ref{Operations}. The 4D-tensor of the source domain in stage one $z_{s}^{1} \in \mathbb{R}^{n_{s} \times f_{s} \times e_{s} \times p_{s}}$ extracted by the feature extractor is first permuted (Fig. \ref{Operations}A(a) $\rightarrow$ Fig. \ref{Operations}A(b)) and then being reshaped (Fig. \ref{Operations}A(b) $\rightarrow$ Fig. \ref{Operations}A(c)) as below:

\begin{equation}
\begin{cases}
\begin{aligned}
&  z_{s}^{1} \in \mathbb{R}^{n_{s} \times f_{s} \times e_{s} \times p_{s}} \xrightarrow{\text{Permute}} \tilde z_{s}^{1} \in \mathbb{R}^{e_{s} \times n_{s} \times f_{s} \times p_{s}}\\
& \tilde z_{s}^{1} \in \mathbb{R}^{e_{s} \times n_{s} \times f_{s} \times p_{s}} \xrightarrow[t_{s} = f_{s} \times p_{s}]{\text{Reshape}} r_{s}^{1} \in \mathbb{R}^{e_{s} \times n_{s} \times t_{s}} \\
\end{aligned}
\end{cases}
\end{equation}
where  $n_{s}$, $f_{s}$, $e_{s}$, and $p_{s}$ denote the samples, filters, electrodes, and feature points of the extracted tensor  $z_{s}^{1}$, $\tilde z_{s}^{1}$ is the 4D-tensor obtained by $z_{s}^{1}$ with dimension permuted from $n_{s} \times f_{s} \times e_{s} \times p_{s}$ to $e_{s} \times n_{s} \times f_{s} \times p_{s}$, characterising the features of each electrode; $r_{s}^{1}$ is the 3D-tensor obtained by $\tilde z_{s}^{1}$ with the incorporation of all the feature points in the last two dimensions, reshaping the dimension from  $e_{s} \times n_{s} \times f_{s} \times p_{s}$ to $e_{s} \times n_{s} \times t_{s}$, $t_{s}$ denotes the number of feature points from all the filters of the extracted tensor $z_{s}^{1}$ from source domain in ``adaptation stage one".

As shown in Fig. \ref{Operations}B, with the obtained 3D-tensor from the source domains, the ``Expand" operation is then conducted on each tensor, thus enabling the broadcast mechanisms and dot products among multi-dimensional tensors:
\begin{equation}
r_{s(i, j, k)}^{1} \in \mathbb{R}^{ e_{s} \times n_{s} \times t_{s}} \xrightarrow{\text{Expand}} \tilde r_{s(l, i, j, k)}^{1} \in \mathbb{R}^{ 1 \times e_{s} \times n_{s} \times t_{s}}
\end{equation}
where $\tilde r_{s(l, i, j, k)}^{1}$ is the expanded 4D-tensor of source domain obtained by $r_{s(i, j, k)}^{1}$ with dimension expanded from $e_{s} \times n_{s} \times t_{s}$ to $1 \times e_{s} \times n_{s} \times t_{s}$.

Similar to the $z_{s}^{1}$, ``Permute", ``Reshape", and ``Expand" transformation operations are conducted on the 4D-tensor of bridging domain $z_{g} \in \mathbb{R}^{n_{g} \times f_{g} \times e_{g} \times p_{g}}$:

\begin{equation}
\begin{cases}
\begin{aligned}
& z_{g} \in \mathbb{R}^{n_{g} \times f_{g} \times e_{g} \times p_{g}} \xrightarrow{\text{Permute}} \tilde z_{g} \in \mathbb{R}^{e_{g} \times n_{g} \times f_{g} \times p_{g}} \\
& \tilde z_{g} \in \mathbb{R}^{e_{g} \times n_{g} \times f_{g} \times p_{g}} \xrightarrow[t_{g} = f_{g} \times p_{g}]{\text{Reshape}} r_{g} \in \mathbb{R}^{e_{g} \times n_{g} \times t_{g}} \\
& r_{g(i, j, k)} \in \mathbb{R}^{ e_{g} \times n_{g} \times t_{g}} \xrightarrow{\text{Expand}} \tilde r_{g(l, i, j, k)} \in \mathbb{R}^{ 1 \times e_{g} \times n_{g} \times t_{g}} \\
\end{aligned}
\end{cases}
\end{equation}
where $\tilde z_{g}$ is the permuted 4D-tensor obtained by $z_{g}$ with dimension permuted from $n_{g} \times f_{g} \times e_{g} \times p_{g}$ to $e_{g} \times n_{g} \times f_{g} \times p_{g}$, characterising the features of each electrode, $r_{g}$ is the reshaped 3D-tensor obtained by $\tilde z_{g}$ with the incorporation of all the feature points in the last two dimensions, reshaping the dimension from  $e_{g} \times n_{g} \times f_{g} \times p_{g}$ to $e_{g} \times n_{g} \times t_{g}$, $t_{g}$  denotes the number of feature points from all the filters of the tensor $z_{g}$ from bridging domain, $\tilde r_{g(l, i, j, k)}$ is the expanded 4D-tensor of bridging domain obtained by $r_{g(i, j, k)}$ with dimension expanded from $e_{g} \times n_{g} \times t_{g}$ to $1 \times e_{g} \times n_{g} \times t_{g}$.

\subsubsection{Bridging loss function of source domain}

The bridging loss function of the source domain $L_{s}$ describes the data distribution difference between source domain and bridging domain across sessions and electrodes, consists of two ``adaptation stages" of losses $L_{s}^{1}$ and $L_{s}^{2}$:

\begin{equation}
\label{Equation Ls}
L_{s} = L_{s}^{1} + L_{s}^{2}
\end{equation}
where $L_{s}^{1}$ and $L_{s}^{2}$ denote the bridging loss functions of the source domain in two adaptation stages obtained by deep features $z_{s}^{1}$ and $z_{s}^{2}$, respectively, connected with 3-dimensional fully connected layers ``\textit{3D FC Layer 1}" and ``\textit{3D FC Layer 2}" in the green dotted box in Fig. \ref{Framework}A.

In this paper, the one-to-one distribution distances across electrodes and sessions are obtained via specially designed matrix arithmetic operations, characterising and computing the temporal-electrode data distribution difference of all the deep features with low computational overhead. With all the tensors after ``Permute", ``Reshape", and ``Expand" operations, the bridging loss function can be easily computed via dot product among different tensors.

The bridging loss function $L_{s}^{1}$ of the source domain in ``adaptation stage one" can be expressed below:
\begin{equation}
\begin{aligned}
& L_{s}^{1} = e^{(\varLambda_{r_{s}-\tilde r_{s}}^{1} + \varLambda_{r_{g}-\tilde r_{g}} - 2\varLambda_{r_{s}-\tilde r_{g}}^{1}) / Md(z_s^1+z_g)} \\
& = e^{\frac{\left\| \Delta_{r_{s}-\tilde r_{s}}^1 \left(\Delta_{r_{s} -\tilde r_{s}
}^1\right)^\mathsf{T} + \Delta_{r_{g}-\tilde r_{g}} \left(\Delta_{r_{g}-\tilde r_{g}}\right)^\mathsf{T} - 2\Delta_{r_{s}-\tilde r_{g}}^1 \left(\Delta_{r_{s}-\tilde r_{g}}^1\right)^\mathsf{T} \right\|_1}{Md(z_s^1+z_g)}} \\
& = e^{\frac{\left\|{(r_{s}^1 \ominus \widetilde r_{s}^1) \cdot (r_{s}^1 \ominus \widetilde r_{s}^1)^\mathsf{T} +
(r_{g} \ominus \widetilde r_{g}) \cdot (r_{g} \ominus \widetilde r_{g})^\mathsf{T}  -2(r_{s}^1 \ominus \widetilde r_{g}) \cdot (r_{s}^1 \ominus \widetilde r_{g})^\mathsf{T}}\right\|_1}{Md(z_s^1+z_g)}} \\
\end{aligned}
\end{equation}
where $\varLambda_{r_{s}-\tilde r_{s}}^{1}$, $\varLambda_{r_{g}-\tilde r_{g}}$, and $\varLambda_{r_{s}-\tilde r_{g}}^{1}$ denote the electrode distances across all the feature points (consisting of all the sessions/trials) of intra-source, intra-bridging, and between source and bridging domains in adaptation stage one, $Md(\cdot)$ represents the computation of the median value, $\Delta_{r_{s}-\tilde r_{s}}^1$, $\Delta_{r_{g}-\tilde r_{g}}$, and $\Delta_{r_{s}-\tilde r_{g}}^1$ denote the one-to-one electrode ``Manhattan distances" across all the feature points of intra-source, intra-bridging, and between source and bridging domains in adaptation stage one, $\left\| \cdot \right\|_1$ denotes the ``$\mathcal{L}_{1}$ Norm" of the matrix, $\ominus$ denotes the broadcast operation.

The computation of the electrode distances $\varLambda_{r_{s}-\tilde r_{s}}^{1}$, $\varLambda_{r_{g}-\tilde r_{g}}$, and $\varLambda_{r_{s}-\tilde r_{g}}^{1}$ is the core of the established matrix arithmetic operation, where distances among all the feature points across electrodes are computed using dot products between tensors. 

Let us take the computation of intra-source domain electrode distances in adaptation stage one $\varLambda_{r_{s}-\tilde r_{s}}^{1}$ as an example for detailed analysis. To obtain electrode distances $\varLambda_{r_{s}-\tilde r_{s}}^{1}$, the one-to-one electrode ``Manhattan distances" across all the feature points are first computed using broadcasting mechanisms:
\begin{equation}
\small
\begin{aligned}
& \Delta_{r_{s}-\tilde r_{{s}_{(e_{1}, e_{2}, \text{:}, \text{:})}}}^1 = r_{s_{(e_{1}, \text{:}, \text{:})}}^1 \ominus \tilde r_{s_{(1, e_{2}, \text{:}, \text{:})}}^1 = \\
& {\left[\begin{array}{cccc}
r_{s_{\left(e_1, 1,1\right)}}^1-\widetilde{r}_{s_{\left(1, e_2, 1,1\right)}}^1 & \cdots & r_{s_{\left(e_{1}, 1, t_s\right)}}^1-\widetilde{r}_{s_{\left(1, e_2, 1, t_s\right)}}^1 \\
\vdots & \ddots & \vdots \\
r_{s_{\left(e_1, n_s, 1\right)}}^1-\widetilde{r}_{s_{\left(1, e_2, n_s, 1\right)}}^1 & \cdots & r_{s_{\left(e_1, n_s, t_s\right)}}^1-\widetilde{r}_{s_{\left(1, e_2, n_s, t_s\right)}}^1
\end{array}\right]} \\
&
\end{aligned}
\end{equation}
where $\Delta_{r_{s}-\tilde r_{{s}_{(e_{1}, e_{2}, \text{:}, \text{:})}}}^1$ is the  ``Manhattan distances" of the intra-source domain in adaptation stage one between electrode pairs $e_{1}$ and $e_{2}$ , consisting of $n_{s} \times t_{s}$ distances across all the samples and feature points the electrode pairs.

The electrode distance of  intra-source domain in adaptation stage one $\varLambda_{r_{s} -\tilde r_{s}}^{1}$ can thus be computed by: 
\begin{equation}
\begin{aligned}
&  \varLambda_{r_{s}-\tilde r_{s}}^{1} \\
& = \left\| \Delta_{r_{s}-\widetilde{r}_{s}}^1 \left(\Delta_{r_{s}-\widetilde{r}_{s}}^1\right)^\mathsf{T} \right\|_1 \\
& = \left\| (r_{s}^1 \ominus \widetilde r_{s}^1) \cdot (r_{s}^1 \ominus \widetilde r_{s}^1)^\mathsf{T}\right\|_1 \\
& = (r_{s_{\left(1, 1,1\right)}}^1-\widetilde{r}_{s_{\left(1, 1, 1,1\right)}}^1) \times (r_{s_{\left(1, 1,1\right)}}^1-\widetilde{r}_{s_{\left(1, 1, 1,1\right)}}^1) \\
& \vdots \\
& + (r_{s_{\left(1, 1, t_s\right)}}^1-\widetilde{r}_{s_{\left(1, 1, 1, t_s\right)}}^1) \times (r_{s_{\left(1, 1, t_s\right)}}^1-\widetilde{r}_{s_{\left(1, 1, 1, t_s\right)}}^1) \\
& \vdots \\
& + (r_{s_{\left(e_s, n_s, 1\right)}}^1-\widetilde{r}_{s_{\left(1, e_s, n_s, 1\right)}}^1) \times (r_{s_{\left(e_s, n_s, 1\right)}}^1-\widetilde{r}_{s_{\left(1, e_s, n_s, 1\right)}}^1) \\
& \vdots \\
& + (r_{s_{\left(e_s, n_s, t_s\right)}}^1-\widetilde{r}_{s_{\left(1, e_s, n_s, t_s\right)}}^1) \times (r_{s_{\left(e_s, n_s, t_s\right)}}^1-\widetilde{r}_{s_{\left(1, e_s, n_s, t_s\right)}}^1)
\end{aligned}
\end{equation}

Similar to the computation of $\varLambda_{r_{s}-\tilde r_{s}}^{1}$, the electrode distances of the intra-bridging domain $\varLambda_{r_{g}-\tilde r_{g}}$ and the electrode distances between source and bridging domains in adaptation stage one $\varLambda_{r_{s}-\tilde r_{g}}^{1}$ can be computed by:
\begin{equation}
\small
\begin{aligned}
& \varLambda_{r_{s}-\tilde r_{g}}^{1} \\
& = \left\| \Delta_{r_{s}-\tilde r_{g}}^1 \left(\Delta_{r_{s} -\tilde r_{g}}^1\right)^\mathsf{T} \right\|_1 \\
& = \left\| (r_{s}^1 \ominus \widetilde r_{g}) \cdot (r_{s}^1 \ominus \widetilde r_{g})^\mathsf{T} \right\|_1\\
& = (r_{s_{\left(1, 1,1\right)}}^1-\widetilde{r}_{g_{\left(1, 1, 1,1\right)}}) \times (r_{s_{\left(1, 1, 1\right)}}^1-\widetilde{r}_{g_{\left(1, 1, 1, 1\right)}}) \\
& \vdots \\
& + (r_{s_{\left(e_s, n_s, t_s\right)}}^1-\widetilde{r}_{g_{\left(1, e_g, n_g, t_g\right)}}) \times (r_{s_{\left(e_s, n_s, t_s\right)}}^1-\widetilde{r}_{g_{\left(1, e_g, n_g, t_g\right)}})\\
\\[3mm]
& \varLambda_{r_{g}-\tilde r_{g}}^{1} \\
& = \left\| \Delta_{r_{g}-\tilde r_{g}} \left(\Delta_{r_{g}-\tilde r_{g}}\right)^\mathsf{T} \right\|_1 \\
& = \left\| (r_{g} \ominus \widetilde r_{g}) \cdot (r_{g} \ominus \widetilde r_{g})^\mathsf{T} \right\|_1 \\
& = (r_{g_{\left(1, 1,1\right)}}-\widetilde{r}_{g_{\left(1, 1, 1,1\right)}}) \times (r_{g_{\left(1, 1,1\right)}}-\widetilde{r}_{g_{\left(1, 1, 1,1\right)}}) \\
& \vdots \\
& + (r_{g_{\left(e_g, n_g, t_g\right)}}-\widetilde{r}_{g_{\left(1, e_g, n_g, t_g\right)}}) \times (r_{g_{\left(e_g, n_g, t_g\right)}}-\widetilde{r}_{g_{\left(1, e_g, n_g, t_g\right)}})
\end{aligned}
\end{equation}

Similar to the adaptation stage one, the bridging loss function of the source domain $L_{s}^{2}$ in stage two can be computed by:
\begin{equation}
\begin{aligned}
& L_{s}^{2} = e^{(\varLambda_{r_{s}-\tilde r_{s}}^{2} + \varLambda_{r_{g}-\tilde r_{g}} - 2\varLambda_{r_{s}-\tilde r_{g}}^{2}) / Md(z_s^2+z_g)} \\
& = e^{\frac{\left\| \Delta_{r_{s}-\tilde r_{s}}^2 \left(\Delta_{r_{s} -\tilde r_{s}
}^2\right)^\mathsf{T} + \Delta_{r_{g}-\tilde r_{g}} \left(\Delta_{r_{g}-\tilde r_{g}}\right)^\mathsf{T} - 2\Delta_{r_{s}-\tilde r_{g}}^2 \left(\Delta_{r_{s}-\tilde r_{g}}^2\right)^\mathsf{T} \right\|_1}{Md(z_s^2+z_g)}} \\
& = e^{\frac{\left\|{(r_{s}^2 \ominus \widetilde r_{s}^2) \cdot (r_{s}^2 \ominus \widetilde r_{s}^2)^\mathsf{T} +
(r_{g} \ominus \widetilde r_{g}) \cdot (r_{g} \ominus \widetilde r_{g})^\mathsf{T}  -2(r_{s}^2 \ominus \widetilde r_{g}) \cdot (r_{s}^2 \ominus \widetilde r_{g})^\mathsf{T}}\right\|_1}{Md(z_s^2+z_g)}} \\
\end{aligned}
\end{equation}
where $\varLambda_{r_{s}-\tilde r_{s}}^{2}$ and $\varLambda_{r_{s}-\tilde r_{g}}^{2}$ denote the electrode distances across all the feature points (consisting of all the sessions/trials) of intra-source in adaptation stage two, and between source and bridging domains,  $\Delta_{r_{s}-\tilde r_{s}}^2$ and $\Delta_{r_{s}-\tilde r_{g}}^2$ denote the one-to-one electrode ``Manhattan distances" across all the feature points of the intra-source in adaptation stage two, and between source and bridging domains,  $r_{s}^{2}$ and  $\tilde r_{s}^{2}$ are obtained from $z_{s}^{2}$  via ``Permute", ``Reshape", and ``Expand" operations.

\subsubsection{Bridging loss function of target domain}

The bridging loss function of the target domain $L_{t}$ describes the data distribution difference between target domain and the bridging domain across sessions and electrodes. Similar to the source domain, the bridging loss function of the target domain can be denoted as:

\begin{equation}
\label{Equation Lt}
L_{t} = L_{t}^{1} + L_{t}^{2}
\end{equation}
where $L_{t}^{1}$ and $L_{t}^{2}$ denote the bridging loss functions of the target domain in two adaptation stages obtained by deep features $z_{t}^{1}$ and $z_{t}^{2}$, respectively.

The bridging loss function of the target domain $L_{t}^{1}$ in adaptation stage one can be computed by:

\begin{equation}
\begin{aligned}
& L_{t}^{1} = e^{(\varLambda_{r_{t}-\tilde r_{t}}^{1} + \varLambda_{r_{g}-\tilde r_{g}} - 2\varLambda_{r_{t}-\tilde r_{g}}^{1}) / Md(z_t^1+z_g)} \\
& = e^{\frac{\left\| \Delta_{r_{t}-\tilde r_{t}}^1 \left(\Delta_{r_{t} -\tilde r_{t}
}^1\right)^\mathsf{T} + \Delta_{r_{g}-\tilde r_{g}} \left(\Delta_{r_{g}-\tilde r_{g}}\right)^\mathsf{T} - 2\Delta_{r_{t}-\tilde r_{g}}^1 \left(\Delta_{r_{t}-\tilde r_{g}}^1\right)^\mathsf{T} \right\|_1}{Md(z_t^1+z_g)}} \\
& = e^{\frac{\left\|{(r_{t}^1 \ominus \widetilde r_{t}^1) \cdot (r_{t}^1 \ominus \widetilde r_{t}^1)^\mathsf{T} +
(r_{g} \ominus \widetilde r_{g}) \cdot (r_{g} \ominus \widetilde r_{g})^\mathsf{T}  -2(r_{t}^1 \ominus \widetilde r_{g}) \cdot (r_{t}^1 \ominus \widetilde r_{g})^\mathsf{T}}\right\|_1}{Md(z_t^1+z_g)}} \\
\end{aligned}
\end{equation}
where $\varLambda_{r_{t}-\tilde r_{t}}^{1}$ and $\varLambda_{r_{t}-\tilde r_{g}}^{1}$ denote the electrode distances across all the feature points (consisting of all the sessions/trials) of intra-source, and between source and bridging domains in adaptation stage one,  $\Delta_{r_{t}-\tilde r_{t}}^1$ and $\Delta_{r_{t}-\tilde r_{g}}^1$ denote the one-to-one electrode ``Manhattan distances" across all the feature points of the intra-source, and between source and bridging domains in adaptation stage one,  $r_{t}^{1}$ and  $\tilde r_{t}^{1}$ are obtained from $z_{t}^{1}$  via ``Permute", ``Reshape", and ``Expand" operations.

The bridging loss function of the target domain $L_{t}^{2}$ in adaptation stage two can be computed by:

\begin{equation}
\begin{aligned}
& L_{t}^{2} = e^{(\varLambda_{r_{t}-\tilde r_{t}}^{2} + \varLambda_{r_{g}-\tilde r_{g}} - 2\varLambda_{r_{t}-\tilde r_{g}}^{2}) / Md(z_t^2+z_g)} \\
& = e^{\frac{\left\| \Delta_{r_{t}-\tilde r_{t}}^2 \left(\Delta_{r_{t} -\tilde r_{t}
}^2\right)^\mathsf{T} + \Delta_{r_{g}-\tilde r_{g}} \left(\Delta_{r_{g}-\tilde r_{g}}\right)^\mathsf{T} - 2\Delta_{r_{t}-\tilde r_{g}}^2 \left(\Delta_{r_{t}-\tilde r_{g}}^2\right)^\mathsf{T} \right\|_1}{Md(z_t^2+z_g)}} \\
& = e^{\frac{\left\|{(r_{t}^2 \ominus \widetilde r_{t}^2) \cdot (r_{t}^2 \ominus \widetilde r_{t}^2)^\mathsf{T} +
(r_{g} \ominus \widetilde r_{g}) \cdot (r_{g} \ominus \widetilde r_{g})^\mathsf{T}  -2(r_{t}^2 \ominus \widetilde r_{g}) \cdot (r_{t}^2 \ominus \widetilde r_{g})^\mathsf{T}}\right\|_1}{Md(z_t^2+z_g)}} \\
\end{aligned}
\end{equation}
where $\varLambda_{r_{t}-\tilde r_{t}}^{2}$ and $\varLambda_{r_{t}-\tilde r_{g}}^{2}$ denote the electrode distances across all the feature points (consisting of all the sessions/trials) of intra-source, and between source and bridging domains in adaptation stage two,  $\Delta_{r_{t}-\tilde r_{t}}^2$ and $\Delta_{r_{t}-\tilde r_{g}}^2$ denote the one-to-one electrode ``Manhattan distances" across all the feature points of the intra-source, and between source and bridging domains in adaptation stage two,  $r_{t}^{2}$ and  $\tilde r_{t}^{2}$ are obtained from $z_{t}^{2}$  via ``Permute", ``Reshape", and ``Expand" operations.

\subsection{Optimization and testing}

The bridging loss function consists of the data distribution differences from both the source and target domains, measuring the data distribution difference across electrodes of all the sessions under both intra- and inter-subject scenarios. The objective function of the proposed BDAN can be expressed below:

\begin{equation}
\label{Equation J}
\min J = L_{cls} + w_{s} L_{s} + w_{t} L_{t}
\end{equation}
where $L_{cls}$ denotes the classification loss of the source domain, $w_{s}$ and $w_{t}$ represent the tradeoff parameters of the bridging loss functions from the source and target domains, respectively.

The testing stage of the BDAN is shown in Fig. \ref{Framework}B, where the data distribution difference across sessions and electrodes is automatically \cite{Aa2023AdaptiveFuzzyControl} minimised using the trained optimised model \cite{Wang2023MultiObjectiveOptimization}, transferring the known spatio-temporal knowledge from the source to the target domain.

\begin{table*}
\centering
\caption{Classification Accuracy of Proposed BDAN and Comparison Methods}
\label{Comparison_results}
\renewcommand\arraystretch{1.2}
\tabcolsep=0.40cm
\begin{tabular}{c|c|@{\hspace{8pt}}c@{\hspace{12pt}}cccccc@{\hspace{12pt}}c} 
\hline\hline
\textbf{Dataset} & \textbf{STS Task} & \textbf{\textbf{DeepConvNet}} & \textbf{EEGNet} & \textbf{DDC} & \textbf{DDAN} & \textbf{DDAF-C} & \textbf{DAWD} & \textbf{BDAN} \\ 
\hline
\multirow{21}{*}{\rotatebox[origin=c]{90}{\textbf{BCIC-III-IVa (2 Classes)}}} 
        & \textit{aa} $\xrightarrow{}$~\textit{al} & 83.93\% & 82.86\% & 85.00\% & 88.21\% & 82.50\% & 89.64\% & \textbf{90.71\%} \\
        & \textit{aa} $\xrightarrow{}$~\textit{av} & 68.21\% & 68.93\% & 70.00\% & 70.71\% & 66.07\% & 67.50\% & \textbf{75.00\%} \\
        & \textit{aa} $\xrightarrow{}$~\textit{aw} & 80.36\% & 81.43\% & 81.79\% & 83.21\% & 80.00\% & \textbf{85.00\%} & 75.00\% \\
        & \textit{aa} $\xrightarrow{}$~\textit{ay} & 75.71\% & 75.71\% & 76.07\% & 77.50\% & 83.93\% & 77.86\% & \textbf{87.14\%} \\ 
        & \textit{al} $\xrightarrow{}$~\textit{aa} & 75.00\% & 76.43\% & 76.79\% & 77.86\% & 78.93\% & 75.36\% & \textbf{82.50\%} \\
        & \textit{al} $\xrightarrow{}$~\textit{av} & 57.14\% & 58.93\% & 60.71\% & 63.57\% & 65.71\% & 64.29\% & \textbf{77.50\%} \\
        & \textit{al} $\xrightarrow{}$~\textit{aw} & 83.21\% & 85.71\% & 85.71\% & \textbf{87.86\%} & 84.64\% & 86.43\% & 87.14\% \\
        & \textit{al} $\xrightarrow{}$~\textit{ay} & 78.57\% & 79.64\% & 80.36\% & 81.43\% & 83.57\% & 83.57\% & \textbf{89.29\%} \\
        & \textit{av} $\xrightarrow{}$~\textit{aa} & 70.36\% & 70.71\% & 70.71\% & 72.50\% & 73.93\% & \textbf{77.14\%} & 74.64\% \\
        & \textit{av} $\xrightarrow{}$~\textit{al} & 73.21\% & 74.29\% & 74.64\% & 75.36\% & 75.00\% & 77.86\% & \textbf{84.64\%} \\
        & \textit{av} $\xrightarrow{}$~\textit{aw} & 67.14\% & 67.86\% & 68.57\% & 71.07\% & 73.21\% & 71.79\% & \textbf{87.86\%} \\
        & \textit{av} $\xrightarrow{}$~\textit{ay} & 71.79\% & 73.21\% & 74.64\% & 75.71\% & 74.64\% & 81.07\% & \textbf{81.79\%} \\
        & \textit{aw} $\xrightarrow{}$~\textit{aa} & 73.21\% & 74.29\% & 75.00\% & \textbf{77.86\%} & \textbf{77.86\%} & 75.71\% & \textbf{77.86\%} \\
        & \textit{aw} $\xrightarrow{}$~\textit{al} & 86.07\% & 86.07\% & 87.50\% & 89.29\% & 87.50\% & 91.79\% & \textbf{92.50\%} \\
        & \textit{aw} $\xrightarrow{}$~\textit{av} & 64.29\% & 65.00\% & 65.36\% & 66.79\% & 71.43\% & 71.43\% & \textbf{77.14\%} \\
        & \textit{aw} $\xrightarrow{}$~\textit{ay} & 71.79\% & 73.21\% & 73.93\% & 79.29\% & \textbf{84.64\%} & 80.71\% & 75.00\% \\
        & \textit{ay} $\xrightarrow{}$~\textit{aa} & 68.57\% & 67.86\% & 68.93\% & 75.00\% & 80.00\% & 71.79\% & \textbf{81.79\%} \\
        & \textit{ay} $\xrightarrow{}$~\textit{al} & 82.86\% & 84.29\% & 85.71\% & 88.57\% & 92.14\% & 90.36\% & \textbf{95.71\%} \\
        & \textit{ay} $\xrightarrow{}$~\textit{av} & 64.64\% & 66.79\% & 66.79\% & 67.86\% & 68.21\% & 67.14\% & \textbf{81.07\%} \\
        & \textit{ay} $\xrightarrow{}$~\textit{aw} & 68.57\% & 68.93\% & 70.71\% & 73.21\% & 73.21\% & 73.57\% & \textbf{83.21\%} \\
        & \textbf{Average} & 73.23\% & 74.11\% & 74.95\% & 77.14\% & 77.86\% & 78.00\% & \textbf{82.87\%} \\ [0.2ex]
\hline
\multirow{21}{*}{\rotatebox[origin=c]{90}{\textbf{BCIC-IV-2a (4 Classes)}}}
        & \textit{A1} $\xrightarrow{}$ \textit{A2} & 47.86\% & 41.07\% & 45.71\% & 48.21\% & 47.86\% & 47.50\% & \textbf{53.21\%} \\ 
        & \textit{A1} $\xrightarrow{}$ \textit{A3} & 59.64\% & 56.43\% & 59.29\% & 58.93\% & 61.79\% & 59.29\% & \textbf{67.86\%} \\
        & \textit{A1} $\xrightarrow{}$ \textit{A4} & 51.43\% & 48.21\% & 51.07\% & 52.86\% & 52.86\% & 53.93\% & \textbf{60.36\%} \\
        & \textit{A1} $\xrightarrow{}$ \textit{A5} & 45.00\% & 40.36\% & 49.29\% & 46.07\% & 45.71\% & 45.36\% & \textbf{52.50\%} \\
        & \textit{A2} $\xrightarrow{}$ \textit{A1} & 50.00\% & 45.36\% & 47.14\% & 50.00\% & 50.00\% & 48.57\% & \textbf{56.79\%} \\ 
        & \textit{A2} $\xrightarrow{}$ \textit{A3} & 46.79\% & 47.14\% & 48.57\% & 48.57\% & 51.07\% & 46.07\% & \textbf{57.14\%} \\
        & \textit{A2} $\xrightarrow{}$ \textit{A4} & 48.21\% & 42.50\% & 51.43\% & 54.64\% & 53.93\% & 50.00\% & \textbf{55.71\%} \\ 
        & \textit{A2} $\xrightarrow{}$ \textit{A5} & 49.64\% & 54.29\% & 60.71\% & 55.36\% & 52.86\% & 59.29\% & \textbf{64.64\%} \\ 
        & \textit{A3} $\xrightarrow{}$ \textit{A1} & 61.43\% & 54.64\% & 61.07\% & 66.43\% & 65.36\% & 61.79\% & \textbf{71.43\%} \\
        & \textit{A3} $\xrightarrow{}$ \textit{A2} & 46.43\% & 44.64\% & 46.43\% & 50.00\% & 48.21\% & 45.00\% & \textbf{53.21\%} \\
        & \textit{A3} $\xrightarrow{}$ \textit{A4} & 51.79\% & 48.21\% & 52.86\% & 53.93\% & 50.00\% & 53.93\% & \textbf{55.00\%} \\ 
        & \textit{A3} $\xrightarrow{}$ \textit{A5} & 46.79\% & 41.43\% & 45.36\% & 50.71\% & 49.29\% & 47.50\% & \textbf{52.50\%} \\ 
        & \textit{A4} $\xrightarrow{}$ \textit{A1} & 54.64\% & 50.36\% & 53.21\% & 54.29\% & 53.93\% & 53.93\% & \textbf{63.21\%} \\ 
        & \textit{A4} $\xrightarrow{}$ \textit{A2} & 48.21\% & 47.14\% & 50.00\% & 50.36\% & 50.00\% & 49.64\% & \textbf{55.36\%} \\ 
        & \textit{A4} $\xrightarrow{}$ \textit{A3} & 54.64\% & 51.79\% & 54.29\% & 56.07\% & 56.43\% & 52.86\% & \textbf{57.86\%} \\
        & \textit{A4} $\xrightarrow{}$ \textit{A5} & 47.14\% & 42.14\% & 49.64\% & 51.43\% & 52.14\% & 50.71\% & \textbf{59.29\%} \\ 
        & \textit{A5} $\xrightarrow{}$ \textit{A1} & 45.71\% & 47.86\% & 48.57\% & 49.64\% & 48.21\% & 50.00\% & \textbf{58.93\%} \\ 
        & \textit{A5} $\xrightarrow{}$ \textit{A2} & 51.79\% & 54.64\% & 58.21\% & 54.64\% & 56.43\% & 56.07\% & \textbf{66.79\%} \\ 
        & \textit{A5} $\xrightarrow{}$ \textit{A3} & 46.43\% & 43.21\% & 46.43\% & 48.21\% & 48.93\% & 44.29\% & \textbf{55.36\%} \\ 
        & \textit{A5} $\xrightarrow{}$ \textit{A4} & 47.14\% & 52.14\% & 51.07\% & 51.07\% & 55.36\% & 48.93\% & \textbf{58.57\%} \\
        & \textbf{Average} & 50.04\% & 47.68\% & 51.52\% & 52.57\% & 52.52\% & 51.23\% & \textbf{58.79\%} \\ [0.2ex]
\hline\hline
\end{tabular}
\end{table*}

\section{Experiment and Result Analysis}
\label{Experiment and Result Analysis}

In this section, to verify the effectiveness of the proposed BDAN, the STS cross-subject MI classification experiments are conducted using the proposed method and the other state-of-the-art methods. Specifically, experiments are implemented on two open-access EEG-based MI datasets: BCI competition III-IVa dataset (BCIC-III-IVa) \cite{Dornhege2004Dataset} and BCI competition IV-2a dataset (BCIC-IV-2a) \cite{Naeem2006BCI-IV-2a}. With the approval of the University Ethics Committee of Xi'an Jiaotong-Liverpool University, the proposal number of this research is EXT20-01-07 on March 31 2020.\par

\subsection{Dataset description}

\subsubsection{Description of BCIC-III-IVa dataset}
The BCIC-III-IVa consists of MI signals from five healthy subjects: \textit{aa}, \textit{al}, \textit{av}, \textit{aw}, and \textit{ay}. There are 280 MI samples for each subject with two MI classes: right-hand and right-foot. The MI signal contains 118 electrode channels following the 10/20 international electrode system and lasts for 3.5 seconds with signals sampled at 100 Hz. The subjects conducted one kind of MI task based on different visual cues and relaxed for 1.75-2.25 seconds in the collection stage. Bidirectional STS experiments are conducted on all the five subjects and there are $A_5^2=20$ STS tasks in total.\par

\subsubsection{Description of BCIC-IV-2a dataset}
The BCIC-IV-2a consists of MI signals from nine healthy subjects: \textit{A1}-\textit{A9}. There are 288 training MI samples for each subject with four MI tasks: left-hand, right-hand, feet, and tongue. The MI signal contains 22 electrode channels and 3 electrooculogram channels, with signals sampled at 250Hz. In the collection stage, the subjects conducted one kind of MI task for 3.0 seconds based on different visual cues. In this paper, subjects \textit{A1}-\textit{A5} are selected for STS cross-subject MI classification tasks and there are $A_5^2=20$ STS tasks in total. \par

\subsection{Experiment settings}
Comparison experiments and ablation studies are performed on datasets BCIC-III-IVa and BCIC-IV-2a to test the effectiveness of the proposed BDAN. Specifically, the proposed BDAN is compared with two deep learning and four deep domain adaptation algorithms, and seven ablation studies have been conducted to show the effectiveness of each part of our algorithm.

\subsubsection{Comparison experiment} The proposed BDAN is compared with two deep learning algorithms and four deep domain adaptation algorithms: DeepConvNet \cite{Schirrmeister2017DeepConvNet}, EEGNet \cite{Lawhern2018EEGNet}, DDC \cite{Tzeng2014DDC}, DDAN \cite{Hang2019DDAN}, DDAF-C \cite{Zhong2023DDAF}, and DAWD \cite{She2023DAWD}. The comparison algorithms mentioned above are introduced in detail as follows:

\begin{itemize}
    \item \textit{DeepConvNet:} a deep CNN architecture inspired by computer vision with deep filters and small kernel size;
    \item \textit{EEGNet:} a compact CNN architecture consists of depthwise and separable convolutions for EEG-based BCI classification, suitable for both intra-subject and cross-subject classification tasks;
    \item \textit{DDC:} a global domain adaptation method for minimizing the global distribution difference of different domains using MMD, aiming to improve the generalization performance of the cross-subject MI classification model;
    \item \textit{DDAN:} a subdomain adaptation method using class subdomain adaptation to maximize the discrepancy of inter-classes in the source domain;
    \item \textit{DDAF-C:} a global domain adaptation method using correlation alignment loss to align the second-order statistical features between the source and target domains;
    \item \textit{DAWD:} a global domain adaptation method using Wasserstein distance to minimize the discrepancy of inter-subject.
\end{itemize}

\subsubsection{Ablation study} Ablation studies are conducted on the proposed algorithms and introduced in detail as follows:

\begin{itemize}
    \item \textit{BDAN-DeepConvNet:} a bridging domain adaptation method revised from the proposed BDAN with feature extractor of DeepConvNet;
    \item \textit{BDAN-EEGNet:} a bridging domain adaptation method revised from the proposed BDAN with feature extractor of EEGNet;
    \item \textit{BDAN-SDA:} a bridging domain adaptation method revised from the proposed BDAN with only the adaptation of the source domain;
    \item \textit{BDAN-TDA:} a bridging domain adaptation method revised from the proposed BDAN with only the adaptation of the target domain;
    \item \textit{BDAN-ST1:} a bridging domain adaptation method revised from the proposed BDAN with only one adaptation stage;
    \item \textit{BDAN-NGK:} a bridging domain adaptation method revised from the proposed BDAN with no Gaussian perturbation in the bridging domain.
\end{itemize}

In both the comparison and ablation experiments, the raw MI data are screened by a 3.5-second window and resampled at 100 Hz. The learning rate is set to 0.001 and gradually decreases along with the milestone for every 50 epochs. The epoch number, batch size, and random seed are set to 500, 40, and 2024. All of the aforementioned methods are compiled in Python using the Pytorch framework. The hardware parameters of the training platform are 32 Intel(R) Xeon(R) Gold 5218 CPUs, 8 GeForce RTX 2080Ti 11GB GPUs, and 180GB RAM.

\begin{figure*}[t]      
    \centering      
    \includegraphics[width=0.91\linewidth]{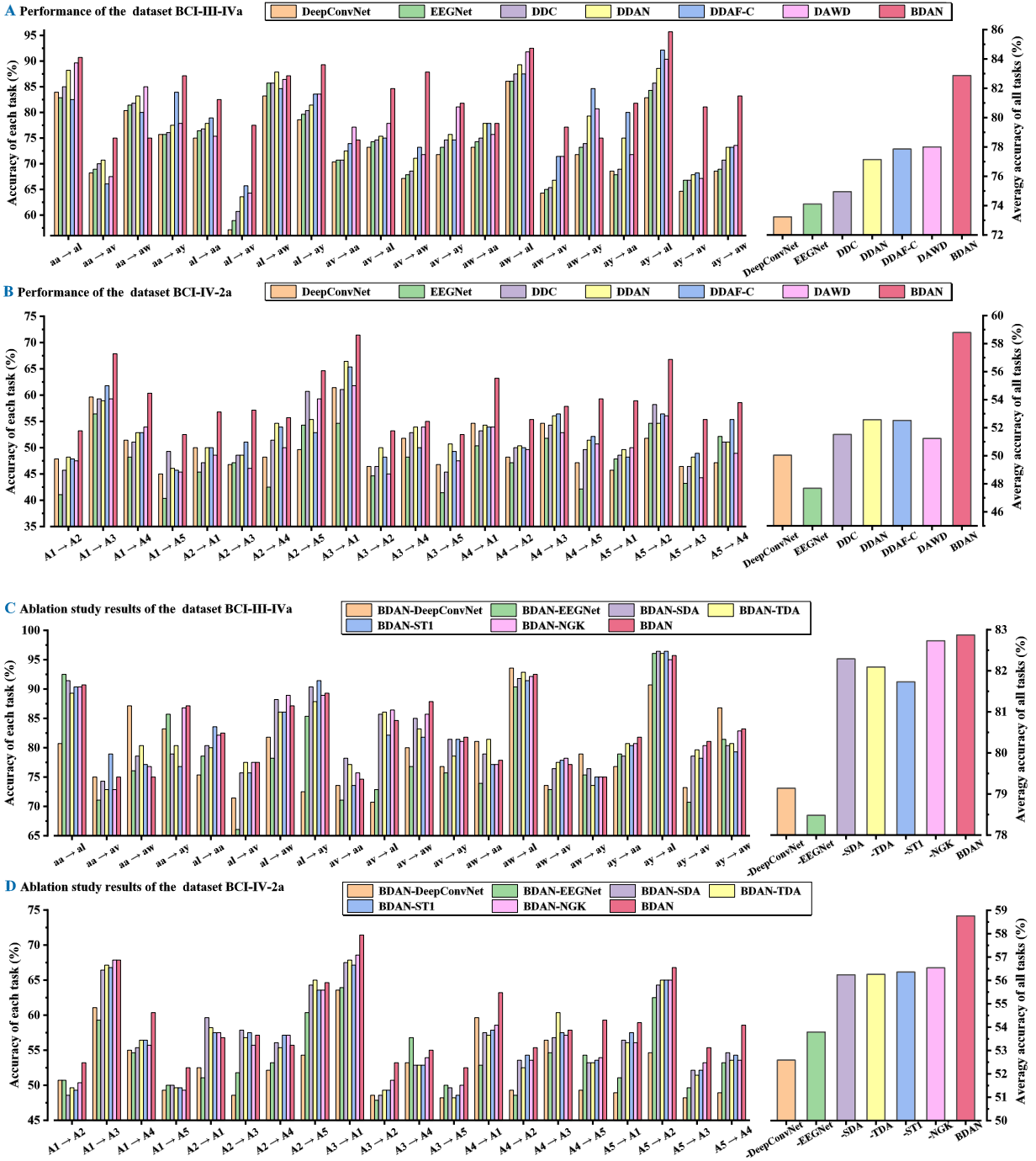}     
    \caption{Performance and results of the proposed BDAN and the other algorithms. \textbf{A-B}, comparison experiments. \textbf{C-D}, ablation studies.} 
    \label{BarChart}
\end{figure*}

\subsection{Comparison results analysis}

To enhance the robustness of the results, 10-fold cross-validation is conducted among all the aforementioned algorithms. Table \ref{Comparison_results} shows the classification results using the proposed BDAN and all the comparison methods on the two datasets, and the best classification accuracy in each task is highlighted in bold. The results of each task and the total average results are plotted as bar charts in Fig. \ref{BarChart}A-B, detailing the performance of our methods.

\begin{table*}
\centering
\caption{Classification Accuracy of all the Ablation Algorithms}
\label{Ablation_results}
\renewcommand\arraystretch{1.2}
\tabcolsep=0.3cm
\begin{tabular}{@{\extracolsep{\fill}}c|c|@{\hspace{1.5pt}}c@{\hspace{0.1pt}}c@{\hspace{2pt}}c@{\hspace{5pt}}ccccc} 
\hline\hline
\textbf{Dataset} & \textbf{STS Task} & \multicolumn{1}{c}{\begin{tabular}[c]{@{}c@{}}\textbf{BDAN-}\\\textbf{EEGNet}\end{tabular}} & \multicolumn{1}{c}{\begin{tabular}[c]{@{}c@{}}\textbf{BDAN-}\\\textbf{DeepConvNet}\end{tabular}}  & \multicolumn{1}{c}{\begin{tabular}[c]{@{}c@{}}\textbf{BDAN-}\\\textbf{SDA}\end{tabular}} & \multicolumn{1}{c}{\begin{tabular}[c]{@{}c@{}}\textbf{BDAN-}\\\textbf{TDA}\end{tabular}} & \multicolumn{1}{c}{\begin{tabular}[c]{@{}c@{}}\textbf{BDAN-}\\\textbf{ST1}\end{tabular}} & \multicolumn{1}{c}{\begin{tabular}[c]{@{}c@{}}\textbf{BDAN-}\\\textbf{NGK}\end{tabular}}  & \textbf{BDAN} \\
\hline
\multirow{21}{*}{\rotatebox[origin=c]{90}{\textbf{BCIC-III-IVa (2 Classes)}}} 
        & \textit{aa} $\xrightarrow{}$~\textit{al} & 80.71\% & \textbf{92.50\%} & 91.43\% & 89.29\% & 90.36\% & 90.36\% & 90.71\% \\
        & \textit{aa} $\xrightarrow{}$~\textit{av} & 75.00\% & 71.07\% & 74.29\% & 72.86\% & \textbf{78.93\%} & 72.86\% & 75.00\% \\
        & \textit{aa} $\xrightarrow{}$~\textit{aw} & \textbf{87.14\%} & 76.07\% & 78.57\% & 80.36\% & 77.14\% & 76.79\% & 75.00\% \\ 
        & \textit{aa} $\xrightarrow{}$~\textit{ay} & 83.21\% & 85.71\% & 78.93\% & 80.36\% & 76.79\% & 86.79\% & \textbf{87.14\%} \\
        & \textit{al} $\xrightarrow{}$~\textit{aa} & 75.36\% & 78.57\% & 80.36\% & 80.00\% & 83.57\% & 82.14\% & \textbf{82.50\%} \\ 
        & \textit{al} $\xrightarrow{}$~\textit{av} & 71.43\% & 66.07\% & 75.71\% & \textbf{77.50\%} & 75.71\% & \textbf{77.50\%} & \textbf{77.50\%} \\
        & \textit{al} $\xrightarrow{}$~\textit{aw} & 81.79\% & 78.21\% & 88.21\% & 86.07\% & 86.07\% & \textbf{88.93\%} & 87.14\% \\
        & \textit{al} $\xrightarrow{}$~\textit{ay} & 72.50\% & 85.36\% & 90.36\% & 87.86\% & \textbf{91.43\%} & 88.93\% & 89.29\% \\
        & \textit{av} $\xrightarrow{}$~\textit{aa} & 73.57\% & 71.07\% & \textbf{78.21\%} & 77.14\% & 73.57\% & 75.71\% & 74.64\% \\
        & \textit{av} $\xrightarrow{}$~\textit{al} & 70.71\% & 72.86\% & 85.71\% & 86.07\% & 82.14\% & \textbf{86.43\%} & 84.64\% \\ 
        & \textit{av} $\xrightarrow{}$~\textit{aw} & 80.00\% & 76.79\% & 85.00\% & 83.21\% & 81.79\% & 85.71\% & \textbf{87.86\%} \\
        & \textit{av} $\xrightarrow{}$~\textit{ay} & 76.79\% & 75.71\% & 81.43\% & 78.57\% & 81.43\% & 81.07\% & \textbf{81.79\%} \\
        & \textit{aw} $\xrightarrow{}$~\textit{aa} & \textbf{81.07\%} & 73.93\% & 78.93\% & 81.43\% & 77.14\% & 77.14\% & 77.86\% \\
        & \textit{aw} $\xrightarrow{}$~\textit{al} & \textbf{93.57\%} & 90.36\% & 91.79\% & 92.86\% & 91.43\% & 92.14\% & 92.50\% \\
        & \textit{aw} $\xrightarrow{}$~\textit{av} & 73.57\% & 72.86\% & 76.43\% & 77.50\% & 77.86\% & \textbf{78.21\%} & 77.14\% \\
        & \textit{aw} $\xrightarrow{}$~\textit{ay} & \textbf{78.93\%} & 75.36\% & 76.43\% & 73.57\% & 75.00\% & 75.00\% & 75.00\% \\
        & \textit{ay} $\xrightarrow{}$~\textit{aa} & 76.79\% & 78.93\% & 78.57\% & 80.71\% & 80.36\% & 80.71\% & \textbf{81.79\%} \\
        & \textit{ay} $\xrightarrow{}$~\textit{al} & 90.71\% & 96.07\% & 96.43\% & 96.07\% & \textbf{96.43\%} & 95.00\% & 95.71\% \\
        & \textit{ay} $\xrightarrow{}$~\textit{av} & 73.21\% & 70.71\% & 78.57\% & 79.64\% & 78.21\% & 80.36\% & \textbf{81.07\%} \\
        & \textit{ay} $\xrightarrow{}$~\textit{aw} & \textbf{86.79\%} & 81.43\% & 80.36\% & 80.71\% & 79.29\% & 82.86\% & 83.21\% \\
        & \textbf{Average} & 79.14\% & 78.48\% & 82.29\% & 82.09\% & 81.73\% & 82.73\% & \textbf{82.87\%} \\
\hline 
\multirow{21}{*}{\rotatebox[origin=c]{90}{\textbf{BCIC-IV-2a (4 Classes)}}}
        & \textit{A1} $\xrightarrow{}$ \textit{A2} & 50.71\% & 50.71\% & 48.57\% & 49.64\% & 49.29\% & 50.36\% & \textbf{53.21\%} \\
        & \textit{A1} $\xrightarrow{}$ \textit{A3} & 61.07\% & 59.29\% & 66.43\% & 67.14\% & 66.79\% & 67.86\% & \textbf{67.86\%} \\
        & \textit{A1} $\xrightarrow{}$ \textit{A4} & 55.00\% & 54.64\% & 55.36\% & 56.43\% & 56.43\% & 55.71\% & \textbf{60.36\%} \\
        & \textit{A1} $\xrightarrow{}$ \textit{A5} & 49.29\% & 50.00\% & 50.00\% & 49.64\% & 49.64\% & 49.29\% & \textbf{52.50\%} \\
        & \textit{A2} $\xrightarrow{}$ \textit{A1} & 52.50\% & 51.07\% & \textbf{59.64\%} & 58.21\% & 57.50\% & 57.50\% & 56.79\% \\
        & \textit{A2} $\xrightarrow{}$ \textit{A3} & 48.57\% & 51.79\% & \textbf{57.86\%} & 56.79\% & 57.50\% & 55.71\% & 57.14\% \\
        & \textit{A2} $\xrightarrow{}$ \textit{A4} & 52.14\% & 53.21\% & 56.07\% & 55.36\% & \textbf{57.14\%} & \textbf{57.14\%} & 55.71\% \\ 
        & \textit{A2} $\xrightarrow{}$ \textit{A5} & 54.29\% & 60.36\% & 64.29\% & \textbf{65.00\%} & 63.57\% & 63.57\% & 64.64\% \\ 
        & \textit{A3} $\xrightarrow{}$ \textit{A1} & 63.57\% & 63.93\% & 67.50\% & 67.86\% & 67.14\% & 68.57\% & \textbf{71.43\%} \\
        & \textit{A3} $\xrightarrow{}$ \textit{A2} & 48.57\% & 47.86\% & 48.57\% & 49.29\% & 49.29\% & 50.71\% & \textbf{53.21\%} \\
        & \textit{A3} $\xrightarrow{}$ \textit{A4} & 53.21\% & \textbf{56.79\%} & 52.86\% & 52.86\% & 52.86\% & 53.93\% & 55.00\% \\
        & \textit{A3} $\xrightarrow{}$ \textit{A5} & 48.21\% & 50.00\% & 49.64\% & 48.21\% & 48.57\% & 50.00\% & \textbf{52.50\%} \\
        & \textit{A4} $\xrightarrow{}$ \textit{A1} & 59.64\% & 52.86\% & 57.50\% & 57.14\% & 57.86\% & 58.57\% & \textbf{63.21\%} \\
        & \textit{A4} $\xrightarrow{}$ \textit{A2} & 49.29\% & 48.57\% & 53.57\% & 52.50\% & 54.29\% & 53.57\% & \textbf{55.36\%} \\
        & \textit{A4} $\xrightarrow{}$ \textit{A3} & 56.43\% & 54.64\% & 56.79\% & 60.36\% & 57.50\% & 57.14\% & \textbf{57.86\%} \\
        & \textit{A4} $\xrightarrow{}$ \textit{A5} & 49.29\% & 54.29\% & 53.21\% & 53.21\% & 53.57\% & 53.93\% & \textbf{59.29\%} \\
        & \textit{A5} $\xrightarrow{}$ \textit{A1} & 48.93\% & 51.07\% & 56.43\% & 56.07\% & 57.50\% & 56.07\% & \textbf{58.93\%} \\
        & \textit{A5} $\xrightarrow{}$ \textit{A2} & 54.64\% & 62.50\% & 64.29\% & 65.00\% & 65.00\% & 65.00\% & \textbf{66.79\%} \\
        & \textit{A5} $\xrightarrow{}$ \textit{A3} & 48.21\% & 49.64\% & 52.14\% & 51.43\% & 52.14\% & 53.21\% & \textbf{55.36\%} \\
        & \textit{A5} $\xrightarrow{}$ \textit{A4} & 48.93\% & 53.21\% & 54.64\% & 53.57\% & 54.29\% & 53.57\% & \textbf{58.57\%} \\
        & \textbf{Average} & 52.62\% & 53.82\% & 56.27\% & 56.29\% & 56.39\% & 56.57\% & \textbf{58.79\%} \\ [0.2ex]
\hline\hline
\end{tabular}
\end{table*}

It can be observed from the bar chart that the proposed BDAN outperforms the other algorithms in most tasks, with a significant improvement in the STS MI classification performance. From the experimental results in Table \ref{Comparison_results} and Fig. \ref{BarChart}A-B, the following observations can be concluded:

1) Compared with the deep learning algorithms DeepConvNet and EEGNet, the proposed BDAN has better classification accuracy in both two datasets. Therefore, the established bridging domain adaptation can reduce the data distribution difference across sessions and electrodes of inter-subject, increasing the performance of the classification model.

2) Compared with the domain adaptation algorithms DDC, DDAN, DDAF-C, and DAWD, the proposed BDAN increases the classification accuracy in both two datasets. The results indicate the bridging domain adaptation can handle the data distribution difference across sessions and electrodes of inter-subject, and the bridging domain adaptation is more powerful than the global one.

\subsection{Ablation results analysis}

Table \ref{Ablation_results} presents the 10-fold cross-validation results of all the aforementioned ablation algorithms, with the best accuracy highlighted in bold. Fig \ref{BarChart}C and \ref{BarChart}D details the performance of all the methods in each task. From the experimental results in Table \ref{Ablation_results} and the bar charts in Fig \ref{BarChart}, the following observations can be obtained:

\begin{figure*}[t]
      \centering
      \includegraphics[width=1\linewidth]{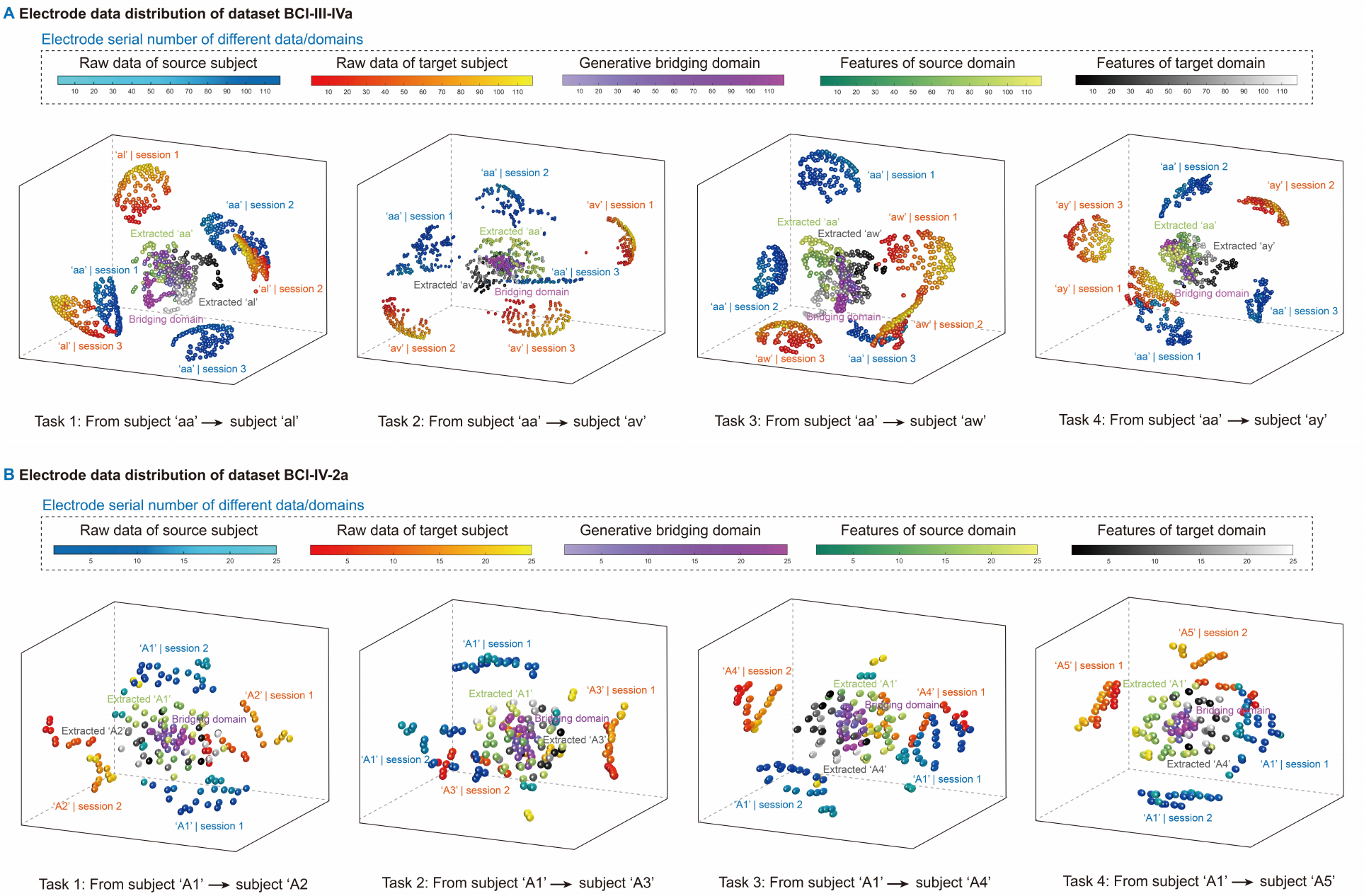}
      \caption{Electrode data distribution visualization of the raw MI data and the deep features extracted by the proposed BDAN, where each electrode ball represents the data distribution of one electrode in different sessions from different subjects. \textbf{A}, electrode data distribution of dataset BCI-III-IVa. \textbf{B}, electrode data distribution of dataset BCI-IV-2a.}
      \label{Visualization_Temporal}
\end{figure*}

(1) Effectiveness of the established bridging loss functions: compared with the deep learning algorithms consisting of no bridging loss functions (DeepConvNet and EEGNet), the BDAN-EEGNet and BDAN-DeepConvNet have a significant increase in classification performance. Such results demonstrate the robustness and effectiveness of the established bridging loss functions (\textit{Remark 3}: the established bridging loss functions are plug-and-play and can be easily embedded into the other deep learning networks).

(2) Effectiveness of the loss functions in both source and target domains: the performance of the two revised BDAN methods (BDAN-SDA and BDAN-TDA) is better than other domain adaptation methods (DDC, DDAN, DDAF-C, and  DAWD), demonstrating the effectiveness of the loss functions of both source and target domains (\textit{Remark 4}: each part in the established bridging loss functions can increase the performance of the classification model and can be used separately).

(3) Effectiveness of the two adaptation stages: compared with the proposed BDAN, the revised BDAN method with only one adaptation stage BDAN-ST1 decreased the classification performance, indicating the effectiveness of both two adaptation stages of BDAN (\textit{Remark 5}: each part in the established bridging loss functions can increase the performance of the classification model and can be used separately).

(4) Effectiveness of the double-Gaussian perturbation in the bridging domain: compared with the proposed BDAN, the revised BDAN method with no Gaussian perturbation decreased the classification performance, especially in the dataset BCI-IV-2a with low electrode density. Such results demonstrate the effectiveness of the double-Gaussian perturbation in the bridging domain. 

\begin{figure*}[!ht]
      \centering
      \includegraphics[width=1\linewidth]{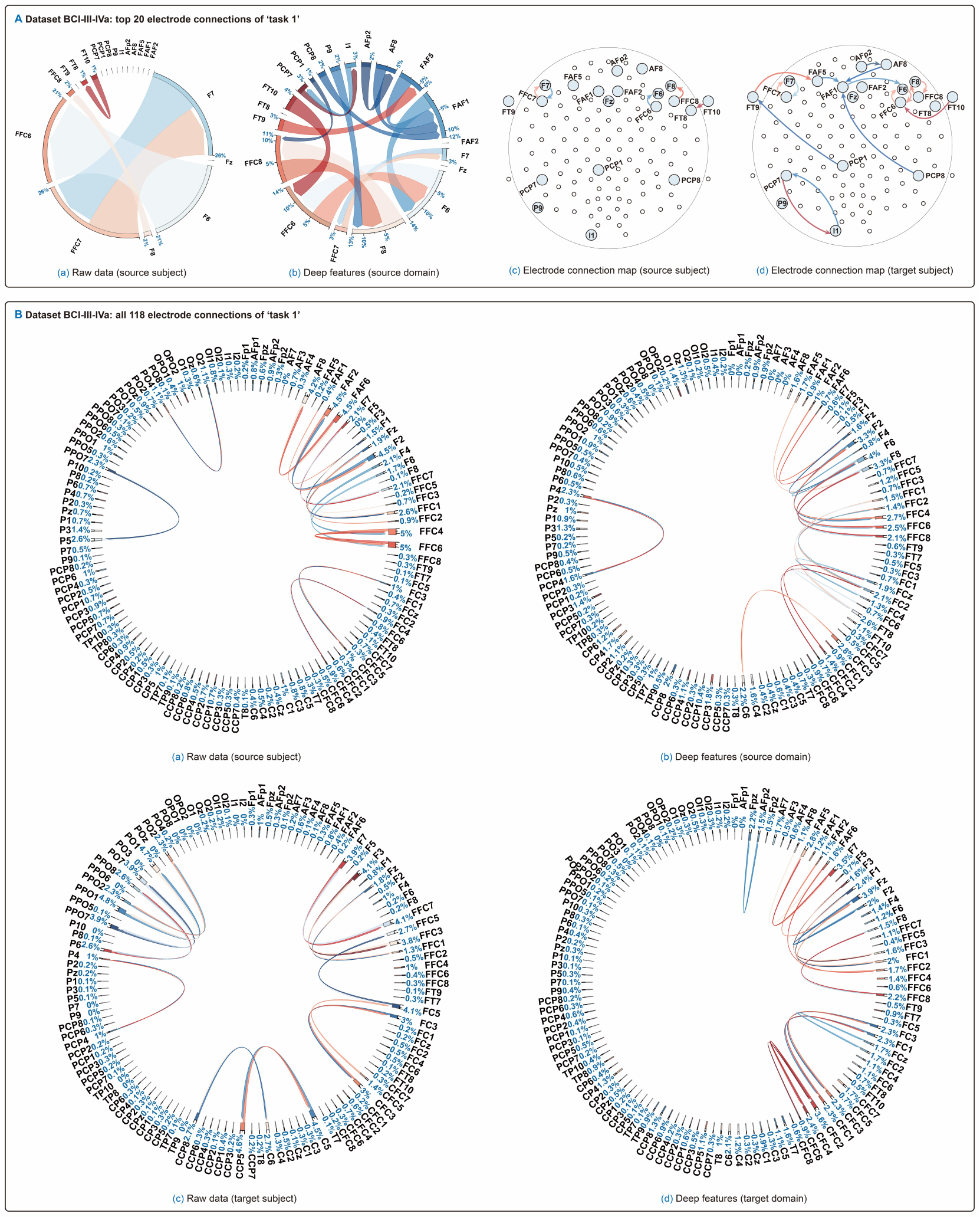}
      \caption{Visualization of the electrode connection strengths of dataset BCI-III-IVa. \textbf{A}, the electrode connection strengths of the top 20 electrodes. \textbf{B}, the top 20\% electrode connection strengths of all the electrodes.}
      \label{Visualization_Electrode}
\end{figure*}

\subsection{Bridging domain adaptation analysis}

To further clarify and demonstrate the performance and effectiveness of the bridging domain adaptation, based on the visualization via 3D-tSNE and chord diagram, detailed analyses are conducted on both temporal and electrode domains between the raw MI data and the deep features.

\subsubsection{Temporal domain analysis}

To illustrate the effectiveness of the BDAN in the temporal domain adaptation, the deep features extracted by the proposed BDAN and the raw EEG data in tasks 1-5 from both two datasets are visualized in Fig. \ref{Visualization_Temporal}. As shown in Fig. \ref{Visualization_Temporal}, each electrode ball represents the data distribution of one electrode from different sessions and domains (indicated by the colour bar at the top of each subfigure). To validly differentiate the data/features from different domains, the data/features are coloured using different gradient colours and coloured texts.

It can be observed that the distribution among electrode balls of the deep features from the inter-subject is much smaller than the raw MI data in each transfer task, with all the sessions and electrodes from both source and target data drawn together after bridging domain adaptation, tightly distributing around the generative bridging domain. Such observation strongly indicates the effectiveness of the established bridging domain adaptation, especially in the aspect of the temporal domain. 

\subsubsection{Electrode domain analysis}

To illustrate the effectiveness of the BDAN in the electrode domain, the connection strengths among electrodes are visualized using chord diagrams as shown in Fig. \ref{Visualization_Electrode}. For a clear demonstration of the relationship among electrodes, we select the dataset BCI-III-IVa with high electrode density (118 electrodes), with the connection strengths visualized separately using the raw MI data and the deep features extracted by BDAN. In Fig. \ref{Visualization_Electrode}, each section of the circle represents one electrode channel, connected via chords with multiple widths (denote the connection strength: thicker chords imply stronger connection strength), indicating the interactions and correlations among electrode channels. For each electrode, the strongest connection among all the connections is selected and plotted to effectively clarify the relationship among electrodes, with the connection strength percentage (compared with the sum of all the connections) illustrated using blue text. To clarify the electrode connections among multiple electrodes with high electrode density, the top 20 electrodes with the strongest connection strengths are selected and visualised in Fig. \ref{Visualization_Electrode}A, with detailed electrode connection maps diagrammatically presented as shown in Fig. \ref{Visualization_Electrode}A(c) and Fig. \ref{Visualization_Electrode}A(d). Furthermore, electrode connections with top 20\% strengths of all 118 electrodes are illustrated in Fig. \ref{Visualization_Electrode}B, further illustrating the effectiveness of the bridging domain adaptation in the electrode domain.

From the chord diagram in Fig. \ref{Visualization_Electrode}, the following observations can be obtained:

a) \textit{For intra-subject scenario}: as shown in Fig. \ref{Visualization_Electrode}A(a) and Fig. \ref{Visualization_Electrode}A(b), compared with the raw data, the number of connections among electrodes (represented by the number of chords) are increased in the deep features after bridging domain adaptation. Detailed electrode connection maps are diagrammatically presented as shown in Fig. \ref{Visualization_Electrode}A(c) and Fig. \ref{Visualization_Electrode}A(d), with all the electrode channels located following the international system. The selected top 20 electrodes are enlarged with the electrode name beside, and the connections between electrodes are indicated using the arrow with the same colour as the corresponding chord. Compared with the electrode connections of raw data, the connections are significantly increased among adjacent electrodes (``F6" and ``FFC6", ``F7" and ``FFC7", ``F8" and ``FFC8", ``AF2" and ``AF8", ``AF8" and ``FAF1", ``FAF5" and ``FAF1", etc.). Obviously, the connection between electrodes after adaptation becomes more consistent with the spatial pattern of EEG signals than raw data, strongly correlated with the pre-defined electrode locations. Such results highlight the effectiveness of the bridging domain adaptation in the electrode domain.

b) \textit{For inter-subject scenario:} as shown in Fig. \ref{Visualization_Electrode}B(a) and \ref{Visualization_Electrode}B(c), the electrode connections of inter-subject among raw MI data are different, with chords appearing between distinct electrode pairs, indicating the electrode data distribution difference of inter-subject. Compared with the connections of the raw MI data, the electrode connections of inter-subject after domain adaptation become similar as shown in Fig. \ref{Visualization_Electrode}B(b) and \ref{Visualization_Electrode}B(d), with strong connections in the same/adjacent electrode pairs between source and target subjects (``FAF2" and ``F2", ``FAF6" and ``F6", ``FAF5" and ``F7/F8", ``FFC8" and ``F8", ``FFC2" and ``F2", ``FC2" and ``CFC2", etc.). Such results demonstrate that the bridging domain adaptation can minimize the data distribution across electrodes of inter-subject.

From the illustrations in aspects a) and b), the proposed bridging domain adaptation increases the connection strengths following the initial electrode arrangement rules, reducing the variability across electrodes of inter-subject. Therefore, the bridging domain adaptation successfully minimises the data distribution difference across electrodes, thus significantly increasing the classification performance of the model.

\section{Conclusion} \label{Conclusion}
Existing research has illustrated the data distribution difference across sessions and electrodes is inevitable in cross-subject motor imagery (MI) classifications, reducing the robustness and performance of the classification model, especially in the single-source to single-target (STS) experiment. In this paper, we investigated and systematically summarised a novel temporal-electrode data distribution difference problem, pointing to the research gap of the existing studies. To tackle the proposed problem, a novel bridging domain adaptation network (BDAN) is proposed in this paper. With the established bridging loss functions, the data distribution across sessions and electrodes is minimized via a specially designed generative bridging domain. The proposed BDAN is tested using 10-fold cross-validation in both comparison and ablation studies and further verified using two novel visualization tools.

Despite the aforementioned advantages of the proposed BDAN, there are still questions involved that deserve to be explored in depth. For example, accompanied by a strong regional pattern connected with the pre-defined electrode locations, the bridging domain adaptation can be further implemented to tackle the data distribution across regions with the help of brain science and imaging techniques. Future works can also be focused on the data distribution difference across hemispheres, especially for motor imagery with regional patterns.

\end{document}